\documentclass[12pt,preprint]{aastex}

\usepackage{graphicx}
\slugcomment{Submitted to The Astrophysical Journal}
\shorttitle{\emph{INTEGRAL} and \emph{RXTE} Observations of Cen A}
\shortauthors{Rothschild et al.}

\begin{document}

\title{\emph{INTEGRAL} and \emph{RXTE} Observations of Centaurus A}

\author{Richard E. Rothschild\altaffilmark{1}, 
J\"orn Wilms\altaffilmark{2},
John Tomsick\altaffilmark{1},
R\"udiger Staubert\altaffilmark{3},
Sara Benlloch\altaffilmark{3},
Werner Collmar\altaffilmark{4},
Grzegorz Madejski\altaffilmark{5},
Sandrine Deluit\altaffilmark{6},
Harish Khandrika\altaffilmark{7}}

\altaffiltext{1}{rrothschild, jtomsick@ucsd.edu, Center for Astrophysics and Space Sciences, University of California,
    San Diego, 9500 Gilman Dr., La Jolla, CA 92093-0424 USA}
\altaffiltext{2}{j.wilms@warwick.ac.uk, Department of Physics,
    University of Warwick, Coventry, CV4 7AL UK}
\altaffiltext{3}{staubert, benlloch@astro.uni-tuebingen.de, Institut
    f\"ur Astronomie und Astrophysik, Sand 1, 72076 T\"ubingen, Germany}
\altaffiltext{4}{wec@mpe.mpg.de, Max-Planck-Institut f\"ur
    extraterrestrische Physik, Giessenbachstrasse, 85748 Garching, Germany}
\altaffiltext{5}{madejski@slac.stanford.edu, Stanford Linear
    Accelerator Center, GLAST Group, 2575 Sand Hill Rd., MS 43A, Menlo
    Park, CA, 94025 USA}
\altaffiltext{6}{sandrine.deluit@cesr.fr, Centre d'Etude-Spatiale des
    Rayonnements, 9, Avenue Du Colonel Roche, BP 4346, 31028 Toulouse
    Cedex 4, France}
\altaffiltext{7}{hkhandrika@berkeley.edu, La Jolla High School, 750
    Nautilus, La Jolla, CA, 92037 USA, now at University of California, Berkeley, CA, USA}

\begin{abstract}
  \emph{INTEGRAL} and \emph{RXTE} performed three simultaneous
  observations of the nearby radio galaxy Centaurus A in 2003 March,
  2004 January, and 2004 February with the goals of investigating the
  geometry and emission processes via the spectral/temporal
  variability of the X-ray/low energy gamma ray flux, and
  intercalibration of the \emph{INTEGRAL} instruments with respect to
  those on \emph{RXTE}. Cen~A was detected by both sets of instruments
  from 3--240 keV. When combined with earlier archival \emph{RXTE}
  results, we find the power law continuum flux and the line-of-sight
  column depth varied independently by 60\% between 2000 January and
  2003 March. Including the three archival \emph{RXTE} observations,
  the iron line flux was essentially unchanging, and from this we
  conclude that the iron line emitting material is distant from the
  site of the continuum emission, and that the origin of the iron line
  flux is still an open question. Taking X-ray spectral measurements
  from satellite missions since 1970 into account, we discover a
  variability in the column depth between 1.0$\times10^{23} \rm
  cm^{-2}$ and 1.5$\times10^{23} \rm cm^{-2}$ separated by
  approximately 20 years, and suggest that variations in the edge of a
  warped accretion disk viewed nearly edge-on might be the cause. The
  \emph{INTEGRAL} OSA~4.2 calibration of JEM-X, ISGRI, and SPI yields
  power law indices consistent with the \emph{RXTE} PCA and HEXTE
  values, but the indices derived from ISGRI alone are about 0.2
  greater. Significant systematics are the limiting factor for
  \emph{INTEGRAL} spectral parameter determination.

\end{abstract}


\section{Introduction}

At a distance of $\sim$3.5 Mpc \citep{Hui93}, the radio galaxy
Centaurus~A is one of the nearest and brightest active galactic nuclei
(AGN).  Since its discovery over three decades ago \citep{Bowyer70}
(Note, however, \citealt{Byrum70}), many X-ray to gamma-ray
instruments have shown it to have non-thermal, power law emission
extending to the MeV range \citep{Baity81,Steinle98}.  \emph{Hubble
Space Telescope} (HST) observations have revealed evidence for a small
($\sim$40 pc), inclined disk of ionized gas (Schreier et al. 1998)
around a $\sim$$10^9$ M$_\odot$ black hole, and perhaps a region near
the black hole evacuated by the jet
\citep{Marconi00}. \citet{Karovska03} suggested that mid-IR
observations resolved the nuclear region of size $\sim$3 pc, and
together this might explain the lack of a hidden broad line region in
Cen~A \citep{Alexander99}.

The power law index of Cen~A below 100 keV has remained at $\sim1.8$
for the last 40 years with the exception of 1972--1973 when
\emph{OSO-7} found the index to be $\sim1.2-1.4$
\citep{Winkler75,Mushotzky76}. The X-ray spectrum does not show any reflection
component or a significant broad iron K$\alpha$-line
\citep{Wozniak98,Rothschild99,Benlloch01}, and yet it is rapidly
variable \citep{Morini89}. This would imply, in contrast to
radio-quiet Seyfert galaxies, that a cold accretion disk extending
down close to the black hole may not be the source of the high energy
radiation and the reprocessor is relatively far away. Furthermore, the
hard X-ray spectra of individual Seyfert~1s show an underlying
continuum which is a power law, with a nearly exponential rolloff with
a folding energy of $\sim$100--300 keV \citep{Johnson97}. Cen A, on
the other hand, does show $\sim$MeV range emission, thereby
indicating it is quite different from the radio-quiet Seyferts.

Current observations of Cen~A are presented in \S~2, and the methods
of analyzing both the \emph{INTEGRAL} and \emph{RXTE} data will be
found in \S~3. In \S~4 we present the results from fitting the
combined instrument data from each mission from the 3 simultaneous
observations and from reanalyzing the 3 previous \emph{RXTE}
observations with HEASOFT release (5.3.1). In \S~5 we use
these results to discuss the emission processes in the nuclear region of
Cen~A, and present our conclusions in \S~6. We present an extensive
analysis of each instrument's response to Cen~A as well as an
inter-calibration of the 2 missions in the Appendix. Included in the
Appendix is also a study of the stability of the PCU2 calibration over
the \emph{RXTE} mission as determined from observations of Cas~A.

\section{Simultaneous \emph{INTEGRAL/RXTE} Observations}

\emph{INTEGRAL} and \emph{RXTE} observed Cen A three times as part of
the \emph{INTEGRAL} AO-1 and \emph{RXTE} AO-7 proposal cycles in March
2003 and early 2004. The \emph{INTEGRAL} observations utilized the
Joint European X-ray Monitor \citep[JEM-X;][]{jemx}, the Integral Soft
Gamma-Ray Imager \citep[ISGRI;][]{isgri} portion of the Imager onBoard
the INTEGRAL Satellite (IBIS), and the SPectrometer on board INTEGRAL
\citep[SPI;][]{spi}. The \emph{RXTE} observations used the
Proportional Counter Array \citep[PCA;][]{Jahoda96} and the High
Energy X-ray Timing Experiment
\citep[HEXTE;][]{Rothschild98}. Table~\ref{tab:obs} gives the details
of these three observations plus the previous three \emph{RXTE}
observations \citep{Benlloch01}. The \emph{INTEGRAL} observations are
essentially continuous due to the $\sim$72 hr elliptical orbit, while
the \emph{RXTE} observations are broken-up due to South Atlantic
Anomaly passages, Earth-occults, and unavoidable slews to other
targets for short periods of time. Consequently, livetimes
for \emph{INTEGRAL} instruments are greater than those for
\emph{RXTE}. The single \emph{RXTE} proportional counter PCU2 has 14
times the background-subtracted counting rate of the JEMX-2 detector,
and HEXTE has 95\% of the ISGRI background-subtracted counting
rate. As seen below, the small PCU2 and HEXTE fields of view
($\sim1^\circ$ FWHM) and low backgrounds result in \emph{RXTE}-derived
spectral parameters with significantly smaller uncertainties than
those derived from \emph{INTEGRAL} spectral data.

\section{Data Analysis Methods}

The analysis of the data from the \emph{INTEGRAL} instruments used the
Off-line Scientific Analysis (OSA) 4.2, and followed the OSA 4.2
procedures outlined in the various instrument analysis user
manuals. Specifically, coded mask deconvolution, as opposed to the
open/closed pixel method, was used.  The \emph{RXTE} instruments
analysis was based upon the Warwick/T\"ubingen/UCSD scripts that make
use of the HEASARC-provided \emph{RXTE} FTOOLS of HEASOFT release
5.3.1.

Due to the variable number of PCUs on at any one time, data for all
observations were available only for PCUs 0 and 2.  We chose to limit
the detector selection further to just PCU2, since PCU0 had lost its
propane veto layer in the Spring of 2000 \citep{Jahoda04}, and
calibration/background estimates for PCU0 were not as mature as those
for PCU2.

An absorbed power law plus narrow iron line spectral model ({\tt
phabs*(power + gauss)}) has been used to represent the flux from Cen~A
for all observations. Since JEM-X did not detect the iron line seen by
the PCA, and since large JEM-X residuals prevented meaningful
estimation of upper limits to the line flux, the line was not included
in the \emph{INTEGRAL} fitting. The following energy ranges were used
in the fitting process: PCU2 (2.5--60 keV), HEXTE (17--240 keV), JEM-X
(3--30 keV), ISGRI (22.5--100 keV), and SPI (20--250 keV).  No
systematic errors have been added to the SPI or HEXTE data. We have
added 0.3\% systematic errors to the PCU2 data in order to obtain
$\chi^2_\nu$ values near 1, since the statistical errors are very much
smaller than the systematic ones. This is consistent with suggested
systematic errors from fitting the Crab \citep{Jahoda04}. (See
Appendix~\ref{sec:pcu2_residuals} for a discussion of additional
specific PCU2 systematic effects that were modeled as part of the
fitting procedure.)  On the other hand, we have had to add systematic
errors to the ISGRI and JEM-X data due to incomplete modeling of the
instrument/background response (see \S~\ref{sec:jemx} \&
\ref{sec:isgri}). The observation-dependent systematic errors added to
JEM-X and ISGRI data are given in
Table~\ref{tab:jemx_isgri_spi_fits}. All errors quoted in this paper
represent 90\% confidence intervals, with the exception of 1$\sigma$
errors for counting rates that were generated by the SHOW RATE command
in XSPEC. Errors on fluxes and equivalent widths for individual
instrument fits were determined using the FLUX and EQW commands within
XSPEC with the ERR option invoked using 500 trials. They rely on the
assumption that the parameter value distribution is multivariate
Gaussian centered on the best-fit parameters, and are only an
approximation in the case that the fit statistic space is not
quadratic\footnote{see
http://heasarc.gsfc.nasa.gov/docs/xanadu/xspec/xspec11/manual/manual.html}.
The PCU2 was taken to be the reference instrument when PCA/HEXTE
spectra were fitted, and when only \emph{INTEGRAL} data were analyzed,
JEM-X spectra provided the reference flux.

The \emph{INTEGRAL} data are comprised of a series of $\sim$2000 s
Science Windows (ScWs) containing the science data. We
investigated the significance of the 20--40 keV flux in the images from
each ScW included in our observation with respect to Cen~A. In Obs.~4,
Revolution 48, ScWs 92--95 had questionable non-detections, assumed to
be due to nearing passage through the radiation belts. These ScWs were
not included in the analysis, while all others were.  We also
identified the ScWs that contained Cen~A in the JEM-X field of view. A
separate list of ScWs than for that of ISGRI and SPI was used for the
JEM-X analysis, due to the smaller JEM-X field of view, and the fact
that the dithering would move Cen~A from the JEM-X field of view. Data
were only accumulated from the inner $4^\circ$ radius region in JEM-X
in order to avoid inclusion of spurious events from the edge of the
detector system.  Consequently, the amount of JEM-X data was less than
that of ISGRI or SPI. No attempt was made to extract PICsIT spectra,
since OSA~4.2 does not provide software for spectral extraction of
PICsIT data. Counting rates and exposure times are given in
Table~\ref{tab:obs}.

The \emph{INTEGRAL} observations were made using the dithering mode of
\emph{INTEGRAL} in which the field of view is stepped around a
selected hexagonal pattern in 2$^\circ$ steps. This is to help
suppress systematic effects associated with background variations in
SPI. The last third of Obs. 4 was taken in the staring mode where the
satellite remained pointed directly at Cen~A.

The \emph{RXTE} data accumulations were restricted to times when Cen~A
was more than $10^\circ$ above the Earth's limb from \emph{RXTE}'s
point of view, when \emph{RXTE} was more than 30 minutes past the
beginning of a South Atlantic Anomaly passage, when the pointing
direction was within $0\fdg01$ of the Cen~A position, and when the
PCU2 and HEXTE high voltages were at their nominal values.

\subsection{\emph{INTEGRAL} Imaging and Spectral Accumulations}

\begin{figure}[t]
\includegraphics[width=6.in]{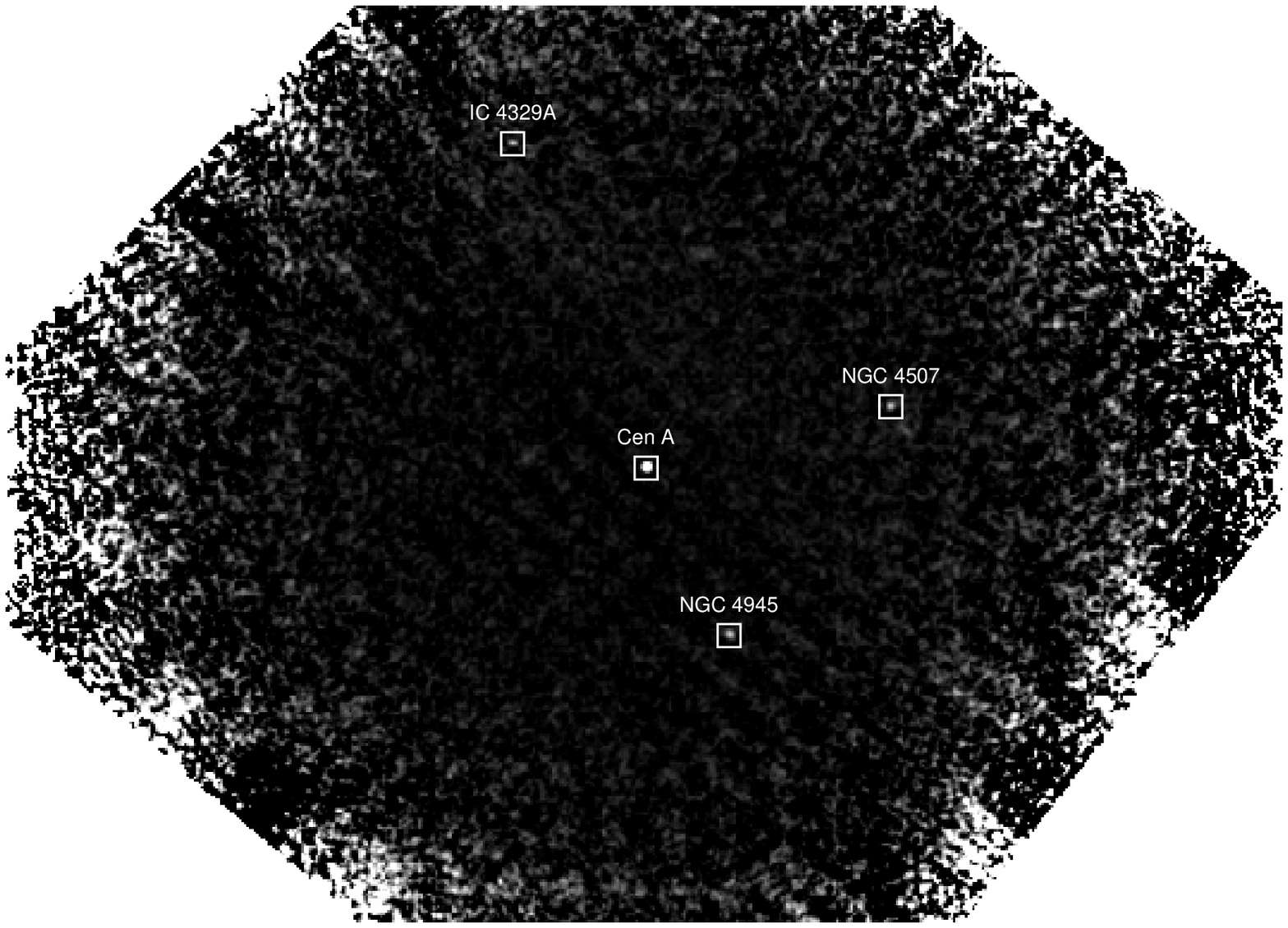}
\caption{The 20--40 keV ISGRI image from Obs. 4, which shows Cen~A,
  NGC~4945, and NGC~4509 as the three sources
  detected.\label{fig:image}}
\end{figure}

Since the ISGRI instrument presented the best view of the multi-source
sky at 20 keV, it was used initially to form images of the field of
view of each observation. Cen A, IC4329A, NGC 4945, and NGC 4509 were the
only sources detected during the Cen~A observations
(Fig.~\ref{fig:image}). The catalog of sources used in subsequent
analyses included only these 3 sources. In addition, FLAG and
SEL\_FLAG parameters in the catalog were set to 1 to force inclusion
of data from portions of the telemetry where the source was in the
field of view but might not meet the detection level criterion.  In this
manner, spectral data were accumulated from all available telemetry.
This approach is essential to the proper measurement of the flux; 
using only the detections above a given threshold will bias the result to
higher fluxes.

The JEM-X spectral histograms were rebinned from the original 256
channels covering 0-82 keV to 34 energy bins with
$\sim0.5$~keV/channel from 3--8 keV, and $\sim1$~keV/channel from
8--30 keV.  The ISGRI data were grouped into 30 energy bins that were
initially 2 keV/channel at 22.5 keV and increased to 100 keV/channel
for 200 keV and beyond.  The present ISGRI analysis was limited,
however, to 22.5--100 keV due to recommendations from the ISGRI team
(P. Ubertini, private communication) and large residuals above 100 keV
(see Appendix~\ref{sec:isgri}). The SPI data were logarithmically
binned into 20 channels from 20--250 keV for the individual
observations and no SPI channels were ignored in the analysis. The
PCU2 spectral data were not rebinned. The HEXTE data remained at 1
keV/channel up to $\sim$50 keV, were then grouped at 5 keV/channel
until $\sim$160 keV above which the binning was 10 keV/channel.

Cen~A was also in the field of view during observations of NGC~4945
taken just after Obs.~5. We have extracted ISGRI and SPI spectra for
this observation also (Cen~A was always outside the JEM-X field of
view) to determine if they could add to the present results. The lower
statistical quality of these spectra due to the reduced off-target
response did not justify their inclusion, but they did provide insight
into results for sources NGC~4945 and NGC~4509 offset from the target
direction during the present observations (see
Appendix~\ref{sec:150}).

\subsection{\emph{RXTE} Spectral Accumulations}

The six spectral accumulations for both PCA and HEXTE were made using
the Standard Data formats present in all \emph{RXTE} observations,
independent of the specific modes chosen by the observers. The PCA
SkyVLE background model was used for PCA background subtraction, while
the HEXTE realtime off-source observations provided the data for HEXTE
background subtraction. The HEXTE instrument design includes
continuous, automatic gain control which allows for the use of a
single instrument response throughout the mission. The accumulation of
background observations of four overlapping regions of the sky just
beyond the source position minimizes any systematic effects due to
temporal and spatial variations in the HEXTE background measurement
within statistical uncertainties. Consequently, systematic
uncertainties in HEXTE source spectra arise only from those in the
instrument response to X-rays and the HEXTE deadtime calculation. The
effect of the latter can be estimated by the size of the adjustment of
the amount of measured background required in an iterative fitting
process, as described for the PCU2
(Appendix~\ref{sec:pcu2_residuals}). For HEXTE, this yielded about
(0.1$\pm$0.1)\% at the 1$\sigma$ uncertainty level. The HEXTE data
were collected in each cluster separately, and the two off-source
positions for each cluster were checked for confusing source(s) by
comparing rates. Upon finding no evidence for such sources, the two
off-source regions of each cluster were added to form each cluster's
background spectrum. Then both the source and background spectral data
from the two clusters were combined to form a single set of HEXTE
source and background files for analysis.

\section{Spectral Analysis of \emph{INTEGRAL} and \emph{RXTE} Data}

\subsection{\emph{INTEGRAL} Results}

\subsubsection{Fitting JEM-X}\label{sec:jemx}

\begin{figure}[t]
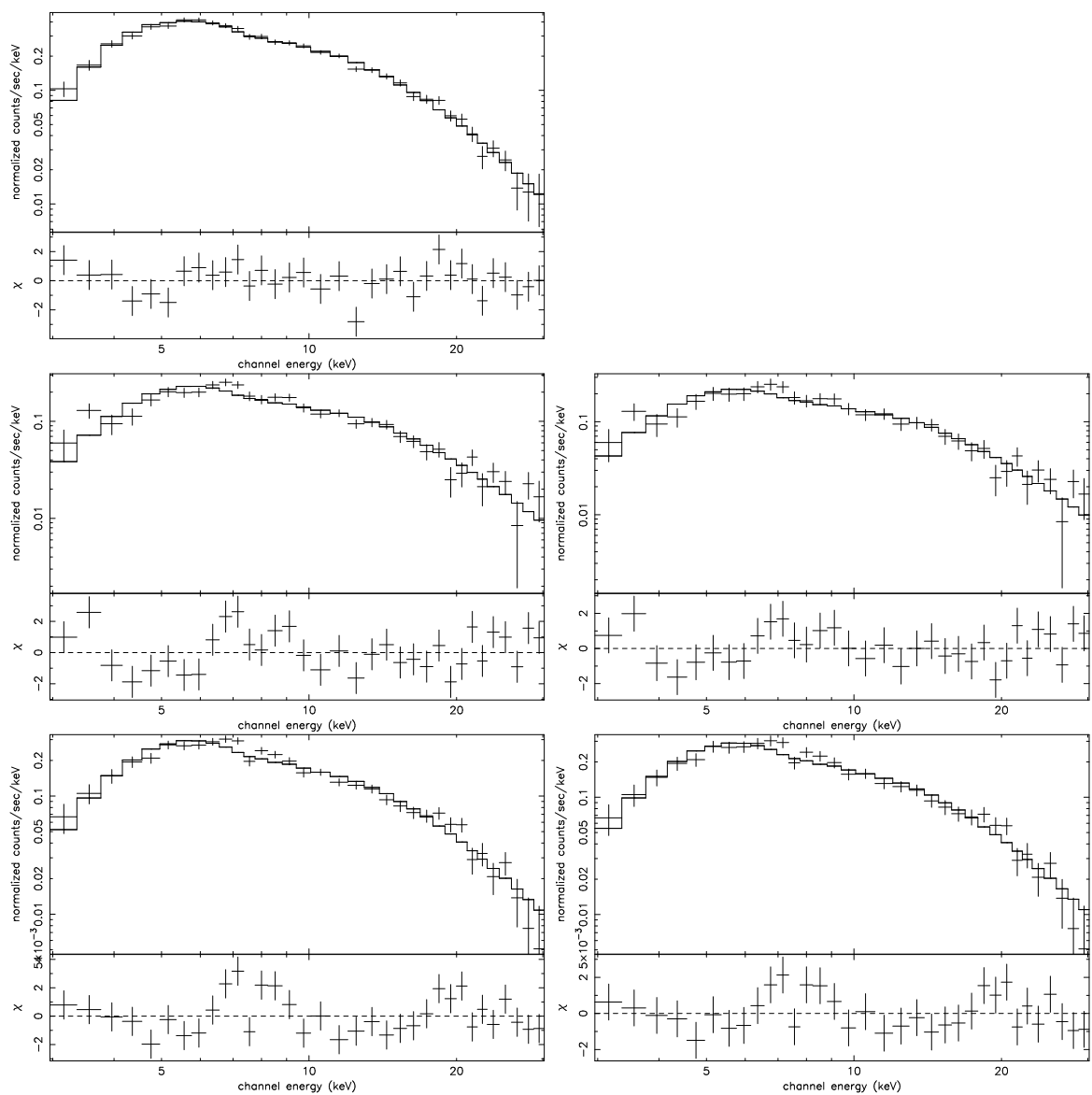

  \includegraphics[angle=270.,width=3.in]{f2a.ps}\\
  \includegraphics[angle=270.,width=3.in]{f2b.ps}
  \includegraphics[angle=270.,width=3.in]{f2c.ps}\\
  \includegraphics[angle=270.,width=3.in]{f2d.ps}
  \includegraphics[angle=270.,width=3.in]{f2e.ps}\\
  \caption{Left: From top to bottom, the JEM-X counts histograms for Obs. 4--6
    without systematic error added (above) and the $\chi$ per energy bin 
  (below) versus energy for fitting to an absorbed power law.
    Right: From top to bottom, fits to the same data with systematic errors of 25\%,
    and 17\% included for Obs. 5, and 6, respectively.  \label{fig:jemx_spectra}}
\end{figure}

The best-fit parameters without additional systematic errors for an
absorbed power law model for the JEM-X observations of Cen~A are given
in Table~\ref{tab:jemx_isgri_spi_fits} and the resulting spectra are
shown on the left hand side of Fig.~\ref{fig:jemx_spectra}. Comparing
the residuals to the fits in the three spectra reveals a broad feature
centered at $\sim$7 keV, which may be the combination of two narrow
lines, in the latter two observations. A third line-like feature is
also seen in the Obs. 6 residuals near 20 keV. Two approaches were
tried to address the systematics in the spectral analysis: 1) the
addition of Gaussian components in the spectral model, and 2) the
addition of systematic error to the spectral data (see
Appendix~\ref{sec:jemx_lines}). Since the first observation has no
strong evidence of these two lines, we speculate that they are
associated with a gain variation with time that is not included in the
JEM-X instrument response calculation. As discussed in
Appendix~\ref{sec:jemx_lines}, the latter prescription of adding
systematic errors to the data to achieve $\chi^2_\nu\approx1$ was
found to be preferable. The results of fitting with systematic
errors added are given in Table~\ref{tab:jemx_isgri_spi_fits} and
shown on the right hand side of Fig.~\ref{fig:jemx_spectra}. The mean
column density $\langle N_{\rm H}
\rangle$=13.8$^{+5.5}_{-4.6}\times10^{22}$ cm$^{-2}$ and the mean
power law index $\langle \Gamma \rangle$=1.80$^{+0.17}_{-0.17}$ for
the 3 JEM-X observations. The 3 individual values of $N_{\rm H}$ and
of $\Gamma$ are consistent with the respective mean values at the 90\%
confidence level. Thus, only variations in the overall flux were
detected by JEM-X.

\subsubsection{Fitting ISGRI}\label{sec:isgri}

\begin{figure}[t]
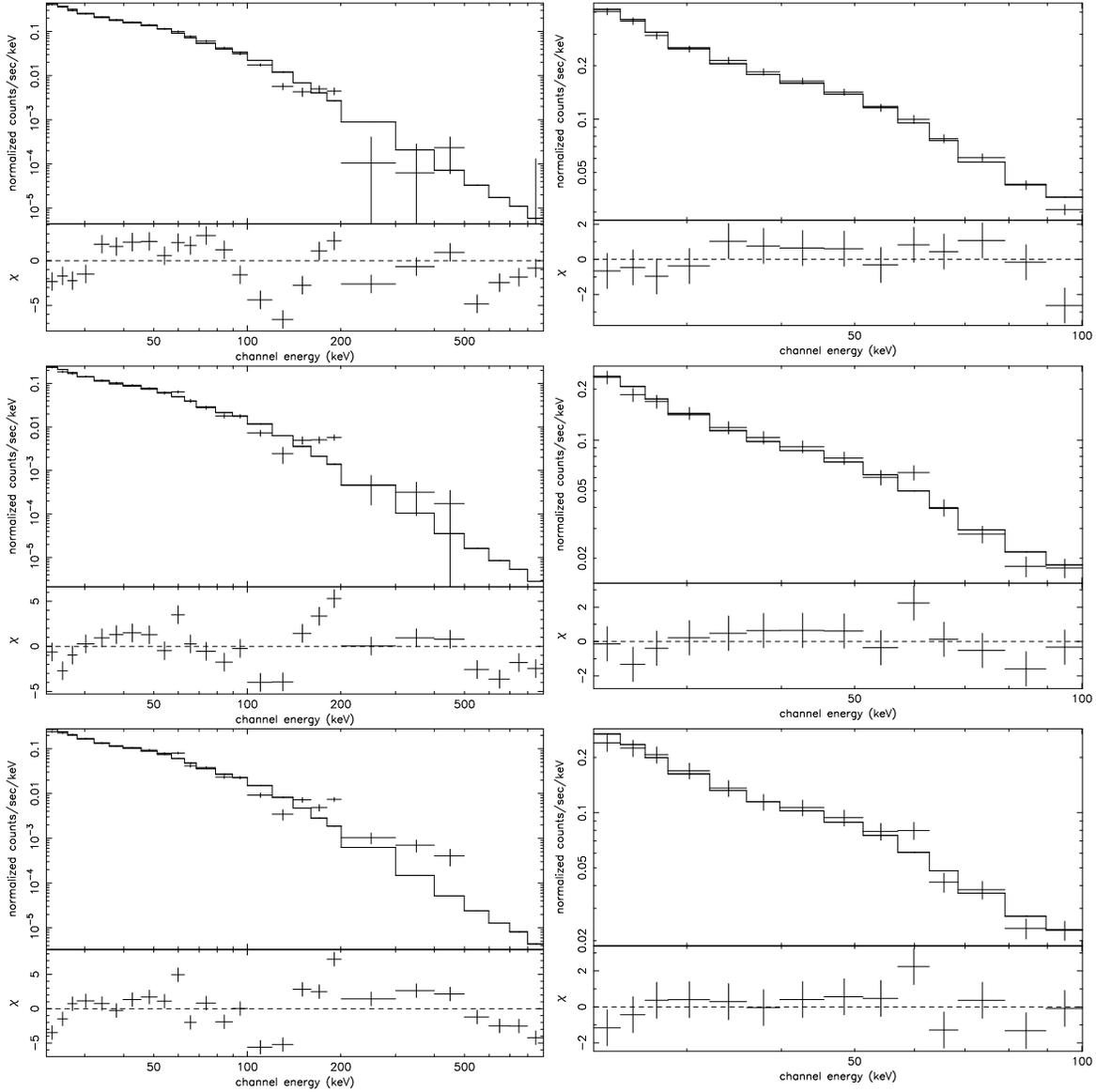

  \includegraphics[angle=270.,width=3.0in]{f3a.ps}
  \includegraphics[angle=270.,width=3.0in]{f3b.ps}\\
  \includegraphics[angle=270.,width=3.0in]{f3c.ps}
  \includegraphics[angle=270.,width=3.0in]{f3d.ps}\\
  \includegraphics[angle=270.,width=3.0in]{f3e.ps}
  \includegraphics[angle=270.,width=3.0in]{f3f.ps}\\
\caption{Left: From top to bottom, the ISGRI counts histograms for
  Obs. 4--6 and best fit models for the full 22.5 keV to 1 MeV energy
  range, and the $\chi$ per energy bin (below) versus energy for
  fitting to an absorbed power law. Note the large negative residual
  above 100 keV as well as the feature at 60 keV and that at the
  lowest energies. Right: From top to bottom, the ISGRI  22.5--100
  keV counts histograms and best fit 22.5--100 keV models
  (above) with 3.5\%, 7.5\%, and 9.5\% systematic errors included,
  respectively, assuming the $N_{\rm H}$ value of the JEM-X fits. The
  $\chi$ per energy bin (below) versus energy for fitting to the
  absorbed power law.
  \label{fig:isgri_spectra}}
\end{figure}

The ISGRI spectra were initially fit to the full 13-400 keV energy
range to determine the extent of significant detections of Cen~A. As
seen in Fig.~\ref{fig:isgri_spectra}~(Left), very significant residuals
to the fit are present below $\sim30$ keV, near 60 keV, and between
100 and 150 keV. These residuals are present in all 3 observations to
some extent. The analysis energy range was then reduced to 22.5--100
keV to concentrate on the high statistical significance portion of the
data, and the fitting was redone. Since the energy range of ISGRI
precluded determination of the line of sight column depth, the JEM-X
value for each observation was
used. Table~\ref{tab:jemx_isgri_spi_fits} gives the best-fit spectral
parameters for the case of no systematic errors added to the data, and
for the case when they were added to achieve $\chi^2_\nu\approx1$.
The trend, however, was to require more systematic errors for the
later observations (3.5\% for Obs.~4, 7.5\% for Obs.~5, and 9.5\% for
Obs.6). This may indicate a time-dependent calibration is required for
ISGRI analysis. Fig.~\ref{fig:isgri_spectra}~(Right) shows the 22.5--100
keV ISGRI spectra with systematics for the 3 observations. The mean
value of the power law index $\langle \Gamma
\rangle$=2.01$^{+0.09}_{-0.09}$, and the 3 individual values of
$\Gamma$ are consistent with this mean value. The
observation-to-observation flux variation seen by JEM-X is also
present in the ISGRI results.

\subsubsection{Fitting SPI}

\begin{figure}[t]
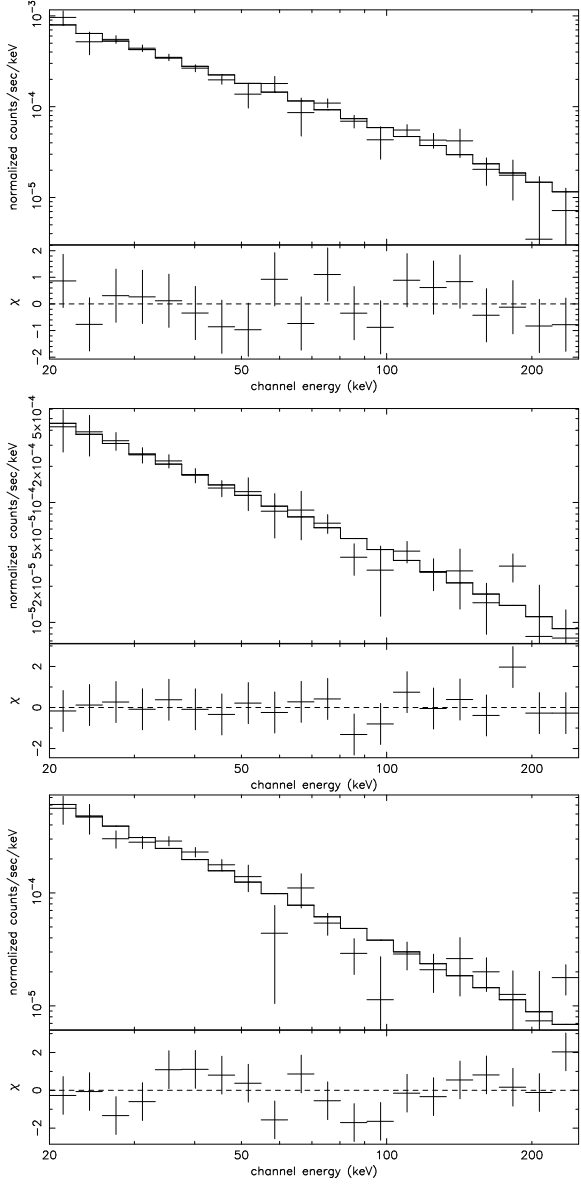

  \includegraphics[angle=270.,width=3.0in]{f4a.ps}\\
  \includegraphics[angle=270.,width=3.0in]{f4b.ps}\\
  \includegraphics[angle=270.,width=3.0in]{f4c.ps}\\
\caption{From top to bottom, SPI counts histograms for Obs. 4--6 (above) and the $\chi$ per energy bin 
  (below) versus energy for fitting to an absorbed power law, assuming
  the $N_{\rm H}$ value of the simultaneous JEM-X fits.\label{fig:spi_spectra}}
\end{figure}

The SPI data were initially accumulated over the 20--600 keV range,
again, to understand the range of detection of Cen~A. This resulted in
our choice to analyze the SPI data from 20--250 keV in 20
logarithmically spaced bins. As was true for ISGRI, the JEM-X value
for the line of sight absorption was used in fitting each SPI
histogram. The 3 SPI spectra are shown in Fig.~\ref{fig:spi_spectra}
and the best-fit parameters are given in
Table~\ref{tab:jemx_isgri_spi_fits}. The range of reduced $\chi^2$ was
0.45$<\chi^2_\nu<$1.10 and thus there was no need for any additional
systematic uncertainties to be added. The mean power law index
$\langle \Gamma \rangle$=1.78$^{+0.17}_{-0.17}$, and the 3 individual
values are consistent with this value. SPI fluxes reflected those seen
by JEM-X.

\subsection{\emph{RXTE} Results}

\subsubsection{Fitting PCU2}
\label{sec:fitting_pcu2}

\begin{figure}[t]
\includegraphics[angle=270.,width=3.in]{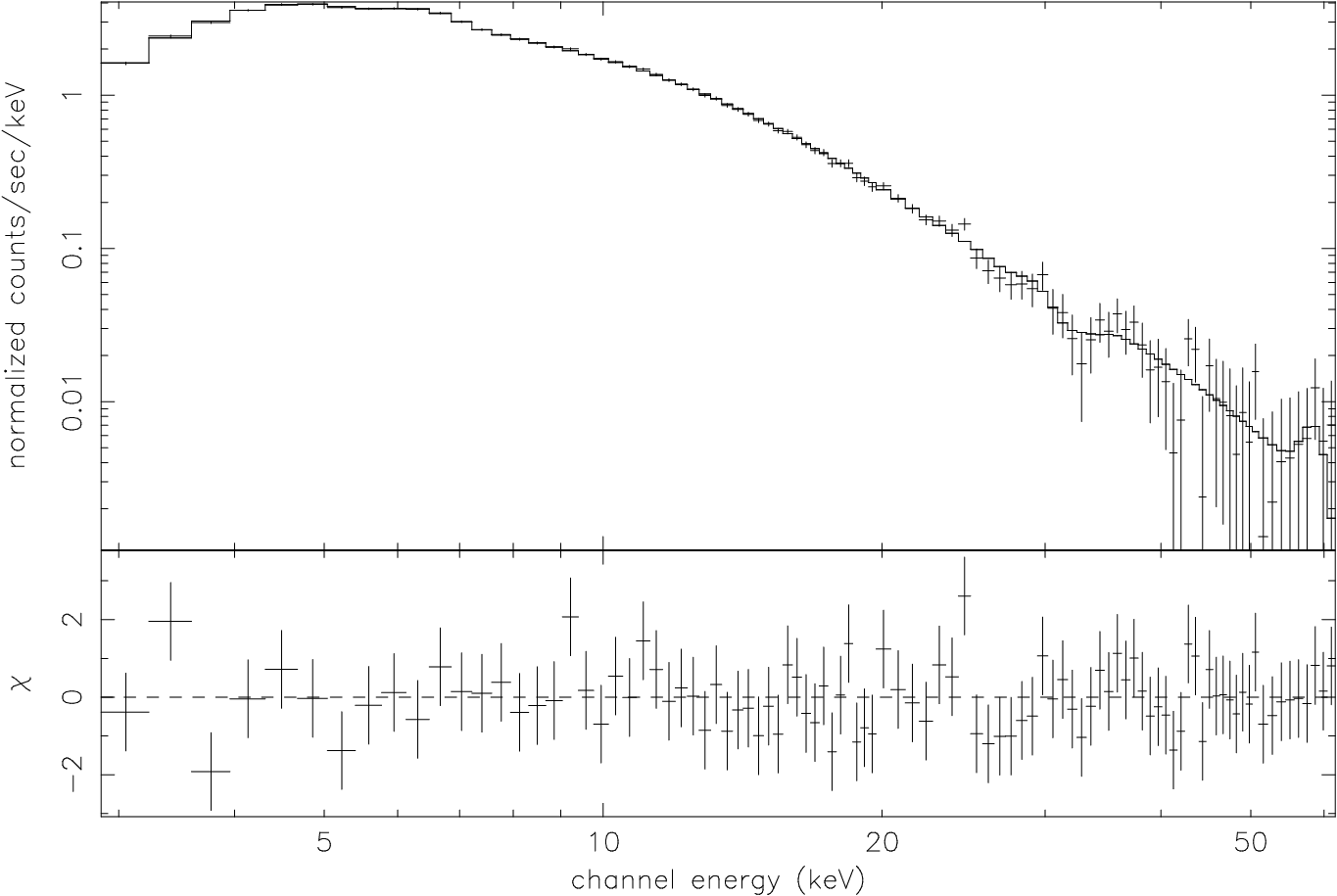}
\includegraphics[angle=270.,width=3.in]{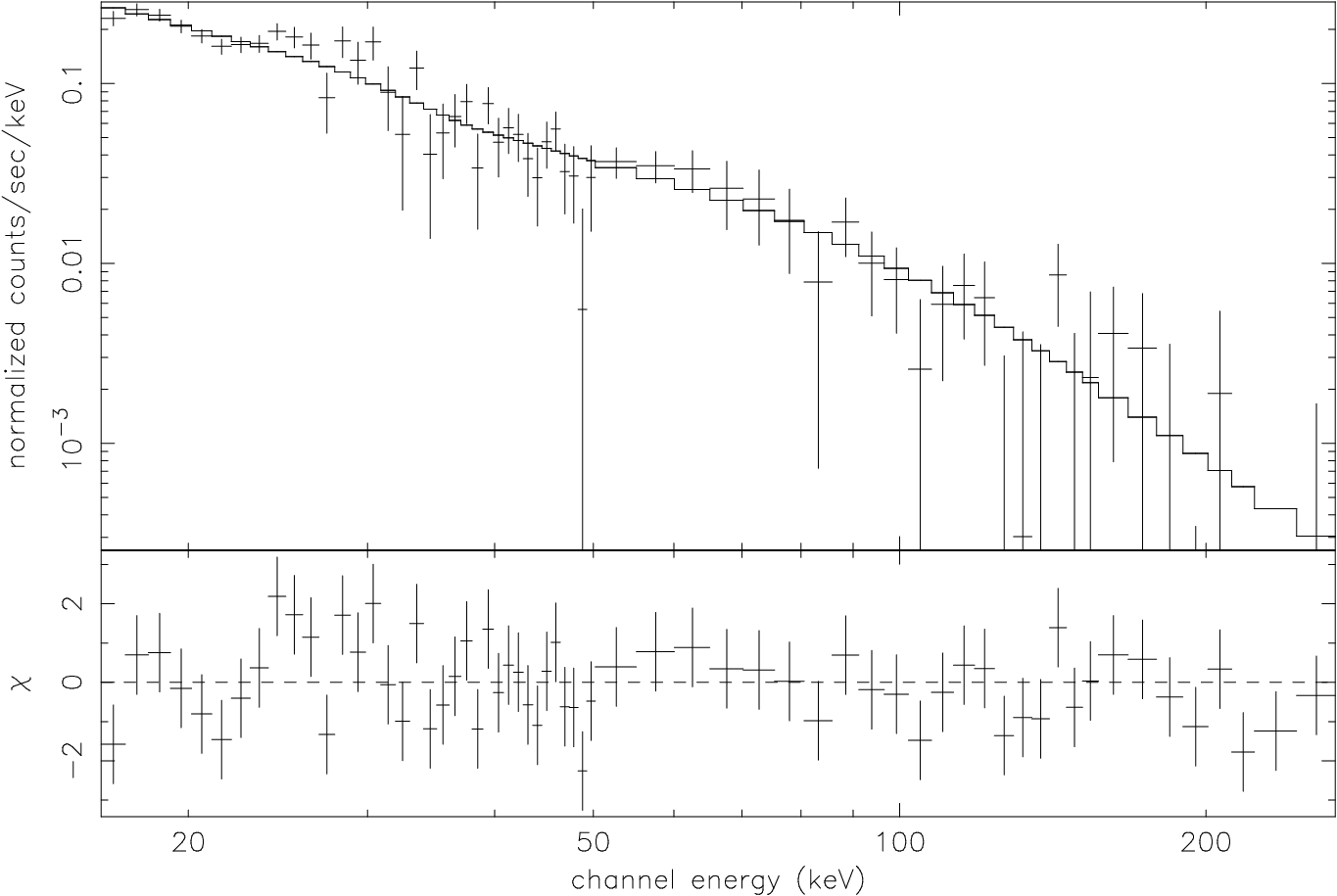}\\
\includegraphics[angle=270.,width=3.in]{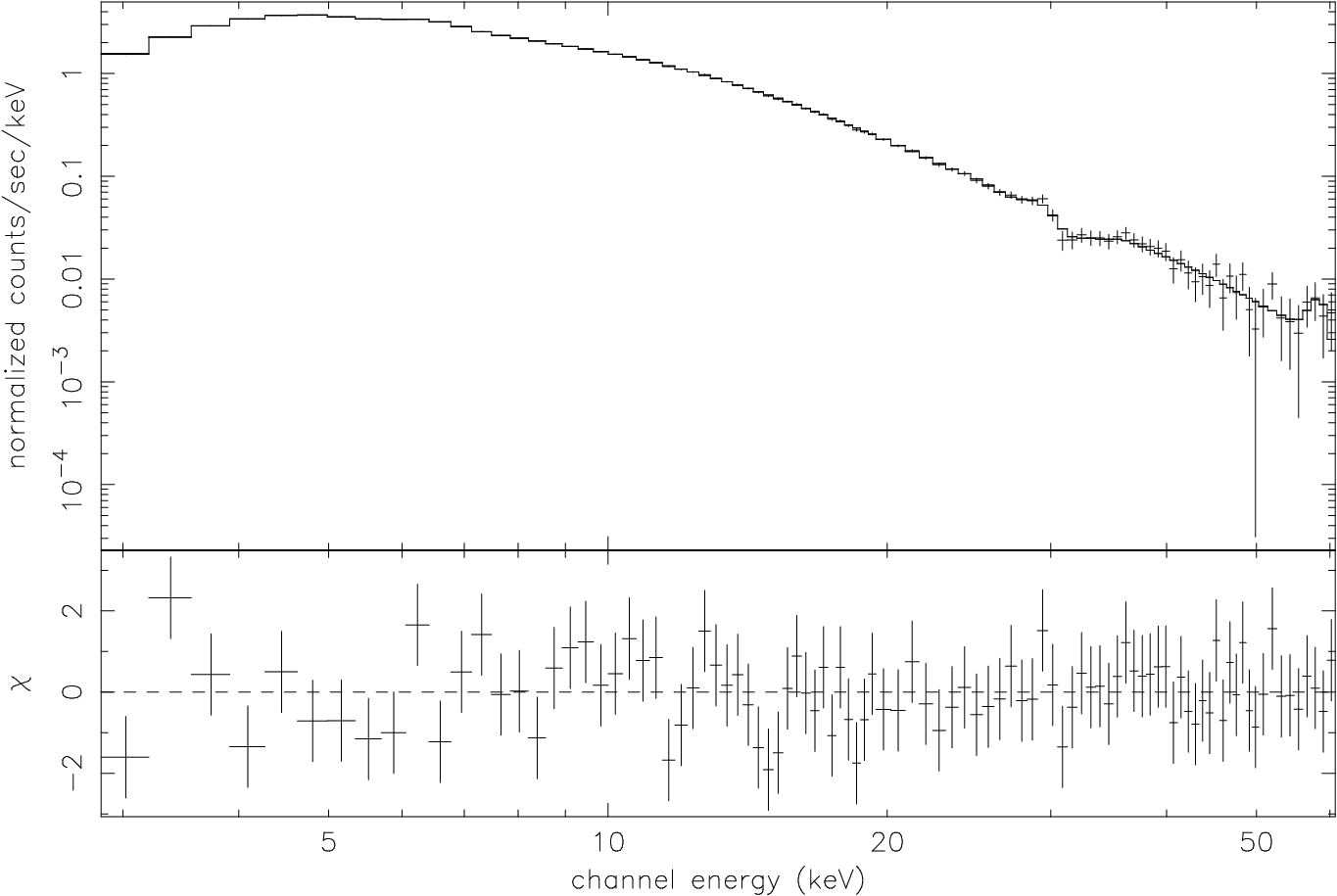}
\includegraphics[angle=270.,width=3.in]{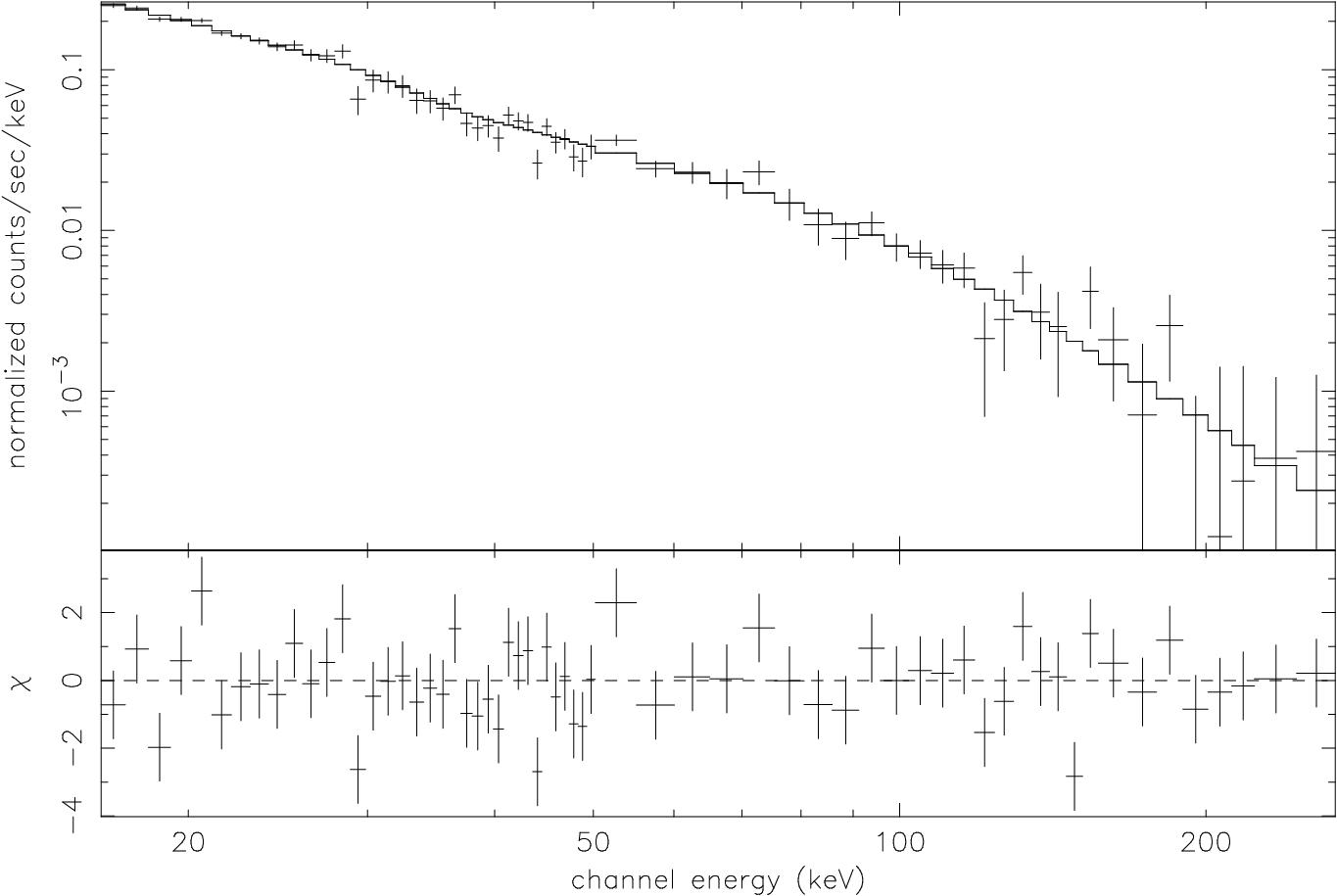}\\
\includegraphics[angle=270.,width=3.in]{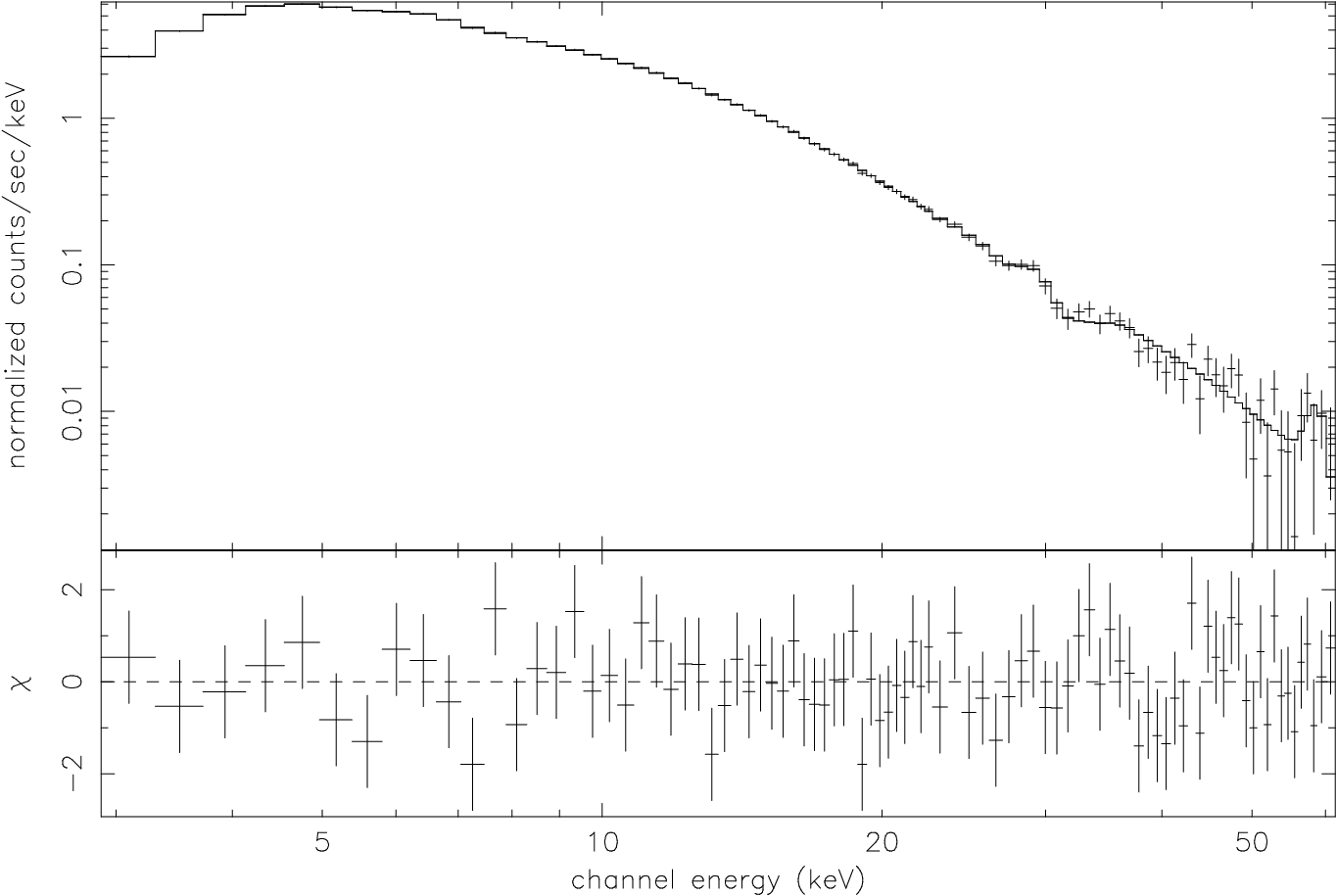}
\includegraphics[angle=270.,width=3.in]{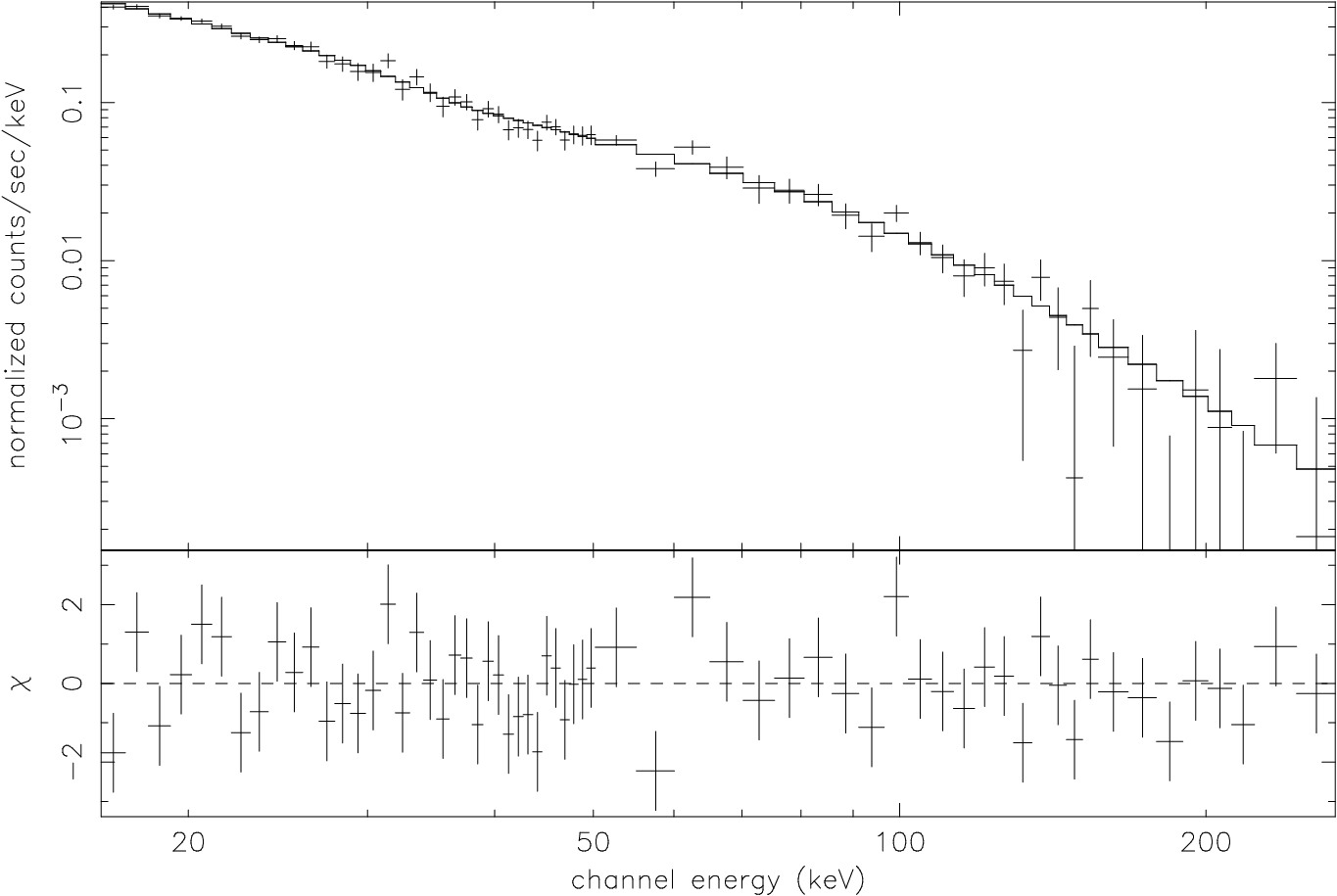}\\
\caption{Left: From top to bottom, PCU2 2.8--60 keV counts histograms for Obs. 1--3
  with instrumental features, background correction, and systematic
  errors included. The residuals to the best-fit histogram are
  displayed as $\chi$ versus energy below each histogram. Right:
  From top to bottom the corresponding HEXTE 17--240 keV counts histograms for Obs. 1--3 with residuals
  displayed below each histogram. \label{fig:pcu2_hexte_first}}
\end{figure}

The residuals to initial fits to PCU2 data and the procedures to
address the systematic effects causing them are discussed in
Appendix~\ref{sec:pcu2_residuals}. By applying these procedures, the
entire 2.8--60 keV energy band of the PCU2 becomes available for
analysis, and at the same time, this expands the overlapping coverage
of PCU2 and HEXTE to 17-60 keV. Table~\ref{tab:pcu2_hexte_fits} gives
the resulting best-fit spectral parameters for the PCU2 observations
of Cen~A, along with the percent corrections made to the background to
enable a best-fit. Corrections to the PCU2 background estimate are at
the few percent level with 1$\sigma$ uncertainties of a few tenths of
a percent. This allows for measurement of continuum parameters and
iron line centroids to $\sim$1\%, and iron line fluxes to
$\sim$5\%. The resulting best-fit spectra are shown in
Figs.~\ref{fig:pcu2_hexte_first}~\&~\ref{fig:pcu2_hexte_last}. In the
first 3 observations, the mean column density and power law index are
$\langle N_{\rm H}\rangle$=9.97$^{+0.27}_{-0.26}\times10^{22}$
cm$^{-2}$ and $\langle\Gamma\rangle$=1.834$^{+0.009}_{-0.032}$, and
the three individual measurements of $N_{\rm H}$ and $\Gamma$ are
consistent with the respective mean values. In the simultaneous
\emph{INTEGRAL/RXTE} observations, $\langle N_{\rm
H}\rangle$=15.86$^{+0.22}_{-0.21}\times10^{22}$ cm$^{-2}$ and
$\langle\Gamma\rangle$=1.829$^{+0.011}_{-0.008}$, and again no
variation in $N_{\rm H}$ or $\Gamma$ is detected for the last 3
observations. When comparing Obs. 1--3 and 4--6 , we find that the
power law indices are consistent with no change, but the column
densities show a 60\% increase. No correlation is seen between the
2--10 keV flux and the column density.

\subsubsection{Fitting HEXTE}

Table~\ref{tab:pcu2_hexte_fits} gives the best fit parameters for the
six HEXTE spectra along with the percentage background adjustments.
We find an average of 0.15\%$\pm$0.12\% adjustment to background, and
are able to achieve 1.01$<\chi^2_\nu<$1.25 when fitting the Cen~A
observational data without application of any additional systematic
errors. The resulting best-fit spectra are shown in
Figs.~\ref{fig:pcu2_hexte_first}~\&~\ref{fig:pcu2_hexte_last}. The
mean power law from the first 3 observations $\langle \Gamma
\rangle$=1.83$^{+0.07}_{-0.07}$ and in the simultaneous
\emph{INTEGRAL/RXTE} observations $\langle \Gamma
\rangle$=1.79$^{+0.02}_{-0.02}$. These values are consistent with
their counterparts from the PCU2 data.

\subsection{Analysis of \emph{INTEGRAL}/\emph{RXTE} Results}
\label{sec:compare}

Comparing the best-fit values from the \emph{RXTE} observations allows
for long term variability to be assessed, and comparing
\emph{INTEGRAL} and \emph{RXTE} results on the last 3 observations is the
basis for evaluating the \emph{INTEGRAL/RXTE} instrument cross
calibration. 

\subsubsection{Instrument Cross-Calibration}

The effective areas of the PCA and HEXTE on \emph{RXTE} have been
extensively calibrated in the laboratory and in orbit with present day
residuals to spectral fitting at the percent level or less. Similarly,
the \emph{INTEGRAL} instruments have had extensive ground
calibrations; however, the absolute effective area of each instrument,
the in-orbit instrument response, and the instrumental background
subtraction technique are still being addressed through regular OSA
releases.

For intercomparison of individual instruments, the \emph{RXTE}/PCU2 is
assumed to provide the ``true'' values of column density and power law
index. The mean PCU2 column density for Obs. 4-6 is
15.86$^{+0.26}_{-0.22}\times10^{22}$ cm$^{-2}$ and the mean best-fit
JEM-X value is 13.8$^{+5.5}_{-4.6}\times10^{22}$ cm$^{-2}$. From this
we see that the JEM-X value is consistent with the PCU2 value at the
90\% confidence level, and the PCU2 is 20 times more sensitive to
column density in this instance. Comparison of the mean power law
indices reveals that JEM-X and SPI agree easily with PCU2 due to their
relatively large uncertainities of 0.17 for indices of 1.80 and 1.78,
respectively, but that the mean ISGRI best fit index is 0.18$\pm$0.09
larger. Averaged over the first 3 observations, HEXTE and PCU2 power
law indices are nearly identical (1.830$^{+0.071}_{-0.075}$
vs. 1.834$^{+0.009}_{-0.032}$), and are a bit farther apart in the
second set of three observations (1.794$^{+0.018}_{-0.018}$
vs. 1.825$^{+0.012}_{-0.010}$). Overall, PCU2 is nearly an order of
magnitude more sensitive to the power law index than the
\emph{INTEGRAL} instruments.

Simultaneous fitting of PCU2 and HEXTE provides for the \emph{RXTE}
best-fit spectral results over the full 2.8--240 keV range and also
yields the HEXTE-to-PCU2 normalization. Simultaneous fitting of JEM-X,
ISGRI, and SPI spectra provides the 3--250 keV \emph{INTEGRAL} result
and normalizations. Table~\ref{tab:integral_rxte_fits} gives the
best-fit Cen~A spectral parameters for \emph{INTEGRAL} and \emph{RXTE}. These
results may differ from those from individual instrument fits, since
they force a single power law to be used, and since column density and
power law index are correlated.

The three \emph{INTEGRAL} instrument effective areas were compared to
\emph{RXTE} by fitting the \emph{INTEGRAL} data with ``frozen''
\emph{RXTE} best-fit spectral parameters, including normalization of the power
law. A variable constant for each \emph{INTEGRAL} instrument's flux was used to
compute their relative normalizations to \emph{RXTE}. In this manner, one
finds the relative normalization of each instrument with respect to
\emph{RXTE}. JEM-X relative to \emph{RXTE} averages
about 90\%, ISGRI about 84\% and SPI about 109\%. HEXTE averaged 92\%
of PCU2 over the 6 observations.

\section{Results on Cen~A}

\subsection{Previous Observations}\label{sec:past_results}

In the years preceeding the present observations, the instruments
on \emph{BeppoSAX} and \emph{RXTE} viewed Cen A five times
\citep{Grandi03} and three times \citep{Benlloch01}, respectively.  In
the energy range above 3 keV, they found the spectrum to be
characterized by an absorbed (N$_H\sim$10$^{23}$ cm$^{-2}$) power law
($\Gamma$=1.80). Simultaneous \emph{BeppoSAX/CGRO} observations,
revealed the spectrum to steepen exponentially (e-folding energy
$\sim$600 keV). The 1991 flight of the Welcome-1 balloon found
evidence for a break, or roll over in the spectrum at 150--200\,keV
\citep{Miyazaki96}, whereas the LEGS balloon (70--500 keV) and
reanalysis of HEAO-1 measurements (2 keV to 2 MeV) of Cen A did not
require a low energy break \citep{Baity81,Gehrels84}.

\emph{Ginga}/OSSE, and \emph{RXTE} found no evidence for a Compton reflection
component implying little cold, Thomson-thick material in the close
vicinity of the AGN or radiation beamed away from the accretion disk
\citep{Wozniak98,Rothschild99,Benlloch01}. The strongest upper limits
to date on the solid angle contributing to the reflection are
$\Omega/2\pi<0.05$ from the combined \emph{RXTE} observations
\citep{Benlloch01}. 

Recently \emph{Chandra} and \emph{XMM-Newton} observations have
allowed the separation of the nuclear component from that of the jet
and galactic flux \citep{Evans04}. These observations resolve the Fe
line and find fluorescent K$_\alpha$ emission from
cold neutral or near-neutral iron with a line width of $\sim$20
eV. This is consistent with emission from material at a large
distance from the site of the hard X-ray emission. The lack of change
in the flux of the Cen~A iron line over 20 years despite significant
changes in the strength of the observed hard X-ray continuum
\citep{Rothschild99}, supports this suggestion that the
absorbing/fluorescing material is at least several parsecs from the
central engine or insensitive to contunuum variations \citep[e.g.][]{Miniutti03}.

\subsection{Cen~A Spectral Results}

With respect to variations in Cen~A, the power law index, the iron
line centroid, and the iron line flux showed no indication of
significant variability over the 6 \emph{RXTE} observations
(Table~\ref{tab:integral_rxte_fits}). The column density, however, was
constant in the pre-2000 observations and again constant in the
post-2000 observations, with a significant 60\% increase between
January 2000 and March 2003 --- (10.0$\pm$0.2)$\times10^{22}$ cm$^{-2}$ to
(15.9$\pm$0.2)$\times10^{22}$ cm$^{-2}$. In conjunction with this
variation in column density, no similar change in iron line flux was
detected. A maximum change of 39\% (4.0$\times10^{-4}$ cm$^{-2}$
s$^{-1}$ to 5.7$\times10^{-4}$ cm$^{-2}$ s$^{-1}$) in the line flux is
the largest allowed within 90\% confidence. The present finding of
constant column density in the 3 earlier \emph{RXTE} observations
compared to the claim of variation by \citet{Benlloch01} can be
attributed to the new calibration (HEASOFT 5.3.1) and to the ability
to analyze PCU2 data to 60 keV.

The PCU2 measured iron line centroid had a mean of
6.42$^{+0.08}_{-0.04}$ keV over the first 3 observations and
6.33$^{+0.07}_{-0.02}$ keV over the second 3. These two values are
consistent with 90\% confidence, and the mean over all 6 observations
is 6.38 keV with a standard deviation of 0.07 keV. The standard
deviation value represents a 1\% measurement that is consistent with
calibration uncertainties and other systematic effects in PCU2
analysis (see Appendix~\ref{sec:casa}). The Fe K$\alpha$ line is
therefore consistent with emission from a neutral medium.

The flux of the iron line had a mean value of
4.43$^{+0.53}_{-0.50}\times10^{-4}$ cm$^{-2}$ s$^{-1}$ for the first 3
observations and 5.13$^{+0.55}_{-0.52}\times10^{-4}$ cm$^{-2}$
s$^{-1}$ in the second three. Thus, no significant flux variability is
detected, as the two values are consistent at the 90\% confidence
level.  The mean of all 6 observations is 4.77$\times10^{-4}$
cm$^{-2}$ s$^{-1}$ with a standard deviation of 0.70$\times10^{-4}$
cm$^{-2}$ s$^{-1}$.  While the standard deviation is about 40\% larger than
the 90\% statistical uncertainties, it only represents about 15\% of the
flux.

The 2--10 keV flux had a factor of 2 range of
(1.69--3.22)$\rm\times10^{-10}ergs\; cm^{-2} s^{-1}$, and no correlation
is seen between flux and power law index, iron line energy, or iron
line flux.

\subsection{Cen~A Lightcurves}

Background subtracted light curves were generated for the full PCU2
energy range with 1024 s time resolution to study temporal variability
over each observation. These light curves are shown in
Fig.~\ref{fig:lc}.  Quarter hour and day-to-day variability of
$\sim$20\% is clearly seen. Fig.~\ref{fig:obs5lc} Top shows the relation
between the On-source, background, and net Cen~A counting rates for
Obs. 5. The ratio of Cen~A to background in PCU2 was $\sim1$.
Fig.~\ref{fig:obs5lc} Bottom gives the light curve of Obs. 2 in more
detail by splitting the observation into first and second halves,
since a large gap in time was inserted into the observation to
accomodate observing another source.

We used the 2.5--60 keV PCA light curves to produce
a power density spectrum (PDS) for the full data set.  We
divided the 16 s time resolution light curves into 1008 s
segments and calculated power spectra for each of the 245
segments.  The 0.001-0.031 Hz PDS includes a total of
246,960 seconds of exposure time.  First, we produced a
Leahy-normalized PDS \citep{Leahy83}, and then we
subtracted-off the Poisson level and re-normalized to
obtain the rms-normalized PDS \citep{Miyamoto91}
shown in Fig.~\ref{fig:rms}.  The PDS is relatively well-described 
($\chi^{2}_\nu$ = 27/20) by a power-law function with a slope
of $\alpha = 2.36\pm 0.15$, and the 0.001-0.03 Hz fractional
rms is 1.59\%$\pm$0.06\%.  There is some evidence for
excess power around 0.018 Hz, but the quality of the power
spectrum is poor above $\sim$0.008 Hz.

\subsection{Search for Spectral Breaks in Cen A}

In order to test for breaks or curvature in the Cen~A continuum, we
fit the three observations with the largest livetime $\times$ PCU2
count rate product to give the best statistical result (Obs. 4, 5, \&
2). In addition to the standard single power law, we tested a broken
power law and a cut-off power law. For Obs. 2 testing of a broken
power law, the fits were insensitive to the break energy and it was
fixed at 100 keV. In each case the iron line and systematics
parameters were essentially unchanged from the single power law case,
and no significant improvement over the single power law was
found. Table~\ref{tab:break} shows the results of this
testing. \textsl{CGRO} found a 1.74$^{+0.05}_{-0.06}$ power law index
from 50 to 150 keV, which then steepened to an index of
2.3$^{+0.1}_{-0.1}$ \citep{Kinzer} when the source flux was
comparable to the present Obs. 4. \textsl{CGRO} found a lesser
steepening ($\Delta \Gamma$=0.24$\pm$0.10) when the flux was lower by a
factor of 0.6, and comparable to Obs. 2. While \textsl{RXTE} yields
results consistent with those of \textsl{CGRO}, the \textsl{RXTE}
spectra do not require the presence of a break.

All 6 HEXTE data sets were summed together to give the highest
statistical sensitivity at high energies, and tested again for a
break. Nothing significant was found. The summed HEXTE spectrum is
shown in Fig.~\ref{fig:hexte_all6}. No deviation from a single power
law to 200 keV was detected.

\subsection{Spectra of Three AGN in the Field}

IC4329a, NGC 4945 and NGC 4507 were in the field of view of
ISGRI and SPI during the Cen~A observations, and as a result, data are
available for them. Their 30--70 keV count rates ranged from 0.6--2
c/s, and detailed spectral analysis was not practical. We fit the
Obs. 4 spectra with a power law and used the resulting best-fit
indices to calculate the 20-100 keV flux for each
object. Table~\ref{tab:ngc} gives the rates and fluxes for both
objects for the 3 \emph{INTEGRAL} observations. The errors given are
1$\sigma$. From this we can conclude that NGC~4945 varied up and down
by $\pm$25\%, NGC~4507 declined over the 3 observations by 50\%, and
IC4329a essentially was constant. As presented in
Appendix~\ref{sec:150}, no additional systematic effects --- just
lower statistical significance due to lower counting rates ---
affected these measurements. The fluxes and power law indices are good
indications (within statistical uncertainties) of the emission
from NGC 4945, NGC 4507, and IC4329a at the time of the Cen~A
observations.

\section{Discussion}

Over the last 3 decades, Cen~A has been observed from space by nearly
all X-ray and gamma ray missions. Fig.~\ref{fig:nh} shows the measured
values of the inferred column density, $N_{\rm H}$, since 1975
\cite[see also][]{Risaliti02}. The lower value of
$\sim$10$\times10^{22}$ cm$^{-2}$ is seen to occur twice in this time
period, with the higher value of $\sim$15$\times10^{22}$ cm$^{-2}$
seen the rest of the time.  The first occurrence of the lower value
was detected by only HEAO-1 in 1978, while the second was seen by
\emph{RXTE, Chandra}, and \emph{BeppoSAX}. One can estimate the
duration of the second occurrence of low column depth to be about 3000
days or 8 years. The duration of the first occurrence could also have
lasted this long, but no observations were made in the early '80s. The
transition from low to high absorption, which may have been resolved
in 2002-2004, took about 2 years. The range of high values of $N_{\rm
H}$ seen from $\sim$(13.5 to 17)$\times10^{22}$ cm$^{-2}$ is broader
than the range of lower values, (9.5 to 10.2)$\times10^{22}$
cm$^{-2}$, and this might indicate that the lower values represent the
baseline for judging variations in column depth. The times of
increased absorption could represent dense ($\sim10^{22}$ cm$^{-2}$)
clouds transiting the line of sight, or variable structure in the
outer edges of the obscuring accretion disk or molecular torus. If the
$\sim$9 year duration of the higher level of absorption seen in the
center of Fig.~\ref{fig:nh} represents a cloud, and if, as discussed
by \citet{Wang86}, it resides in the broad line region at 10$^{17}$ cm
from the central object and has velocity of 500--1000 km/s, its
diameter would be $\sim$10$^{17}$ cm --- the size of the entire broad
line region. If we assume a more reasonable cloud of diameter of
10$^{13}$ cm and $N_{\rm H}=5\times10^{22}$ cm$^{-2}$, then its
velocity would be a meager 0.3 km/s and would place the cloud far
beyond the core region. A cloud-based explanation appears to be
untenable.

The second possibility is variable structure in the outer edge of the
disk. This could be characterized as a non-uniform edge structure that
rotated through the line of sight as the outer disk rotated or just
stochastic variations in disk structure. Assuming a 2$\rm \times10^8
M_\odot$ black hole \citep{Silge05}, 20 pc radius accretion disk
\citep{Schreier98}, and Kelperian motion, the velocity of the outer
edge of the disk is $\sim7\times10^6$ cm s$^{-1}$ and the
circumference is $\sim 4\times10^{21}$ cm. A point on the edge will
travel 2$\times10^{15}$ cm in 8 years, or less than a millionth of the
circumference. Thus, the required structure is quite small with
respect to the disk, and is not out of the question. Another possible
explanation is precession of the warped accretion disk
\citep{Schreier98} creating a variable absorption. The lower column
depth would represent the time when the edge of the disk raised or
lowered to allow a more direct view of the emission region, and the
higher values could be associated with the edge of the disk returning
to attenuate the X-ray emission.

While the changes in the column depth are clear and dramatic over the
past 30 years, the behavior of the inferred power law index is less
so. Fig.~\ref{fig:index} displays the power law indices versus time
for the same missions as Fig.~\ref{fig:nh}. While
the index was less than 1.7 during the \emph{OSO-8/HEAO-1} era, the index is
seen to be consistent with 1.8 since 1989 (\emph{Ginga}). The column
depth is not correlated with the power law index, since the 
$N_{\rm H}=10\times10^{22}$ cm$^{-2}$ and $\Gamma<$1.7 condition was not repeated in
the \emph{RXTE/BeppoSAX} era when the 50\% increase in $N_{\rm H}$
occurred. The power law index at that time did not change from
1.83. This fact is further strengthened by the fact that the two values
of $N_{\rm H}$ and the single value of $\Gamma$ were measured by the
same instrument set on \emph{RXTE}, and henceforth possible systematic effects
relating to differing calibrations on different spacecraft are
not a factor. From this we conclude that the primary emission region
producing the power law component is independent of the absorbing region.

We also note that the X-ray telescopes \emph{ASCA, Chandra} and
\emph{XMM-Newton} have larger uncertainties in their determination of
the continuum than the non-imaging missions with energy ranges
extending to higher values. This highlights the importance of
simultaneous broadband X-ray coverage of \emph{XMM-Newton} and
\emph{Chandra} observations of bright sources, such as black hole
transients and accreting X-ray pulsars where detailed knowledge of the
continuum is essential.

No correlation is seen between flux and power law index, and no large
variation in iron line flux is seen since 1984,  while the inferred
absorbing column varied by 60\%. The flux of the power law continuum
varied a factor of two or more during this time with no accompanying
variation in the line flux (Fig.~\ref{fig:index_flux}).  Two possible
sources of the iron line flux are a reflection component
\citep[e.g.,][]{George91} or transmission through the obscuring matter
\citep[e.g.,][]{Miyazaki96}. Both scenarios, however, are not
consistent with the observations. The reflection component is
attractive, since calculations by \cite{Miniutti03} show that the
power law component can exhibit large variations due to the position
of the primary emission above the accretion disk while the iron line
flux variations would be an order of magnitude less. A reflection
component is not required from the spectral analysis
\citep{Benlloch01}, however, and thus the contribution to the observed
iron line flux would be minimal. On the other hand, transmission
through a line-of-sight absorbing medium (outer edge of an accretion
disk or an obscuring torus) would predict that the equivalent width
would vary with column depth with an equivalent width of $\sim$100 eV
for $N_{\rm H}\approx1\times10^{23}$ cm$^{-2}$ \citep{Miyazaki96}.
Obs. 1--3 as well as Obs. 4--6 find a factor of 2 variation in the
equivalent width for constant values of the column depth of 1 and 1.5
$\times10^{23}$ cm$^{-2}$, respectively.

If a distant iron emitting region were illuminated by beamed emission and
the continuum was produced relatively near the black hole, the sparse
sampling of Cen~A would not have been expected to reveal correlated
behavior. This physical separation would easily support the
observation of reduced equivalent widths with increased continuum
flux, as seen in each set of 3 observations by \emph{RXTE} that
are characterized by a single value of the column depth. If the iron
line flux represents a measure of the beamed flux, estimates of the
line flux indicate that the jet luminosity is on the order of the
X-ray continuum luminosity from the Cen~A core. If this hypothesis is correct,
we can infer that the beamed flux variability, on the average, is less
variable than that of the X-ray flux on the timescale of the 6
\emph{RXTE} observations.

\section{Conclusions}

The six \emph{RXTE} observations over the last nine years have shown
that the 3--240 keV Cen~A spectrum can be described by a single
absorbed power law plus an iron emission line. We have measured the
column density to Cen~A to about 1\%, the power law index and iron
line centroid energy to better than 1\%, and the iron line flux to
approximately 10\%. While still systematics dominated, \emph{INTEGRAL}
determines the power law index to a few percent and the column depth
to 10--20\% from simultaneous observations with the last three
\emph{RXTE} observations in 2003 and 2004. We have provided an in depth
comparison of \emph{RXTE} and \emph{INTEGRAL} instruments using the
latest knowledge of the instrument responses and techniques for
addressing systematic errors. Appendix~\ref{sec:50} gives the improved
\emph{INTEGRAL} spectral results using OSA~5.0 analyzed after
submission of this paper.

The mean values of $N_{\rm H}$ and $\Gamma$ resulting from the
spectral analyses of individual instruments on \emph{INTEGRAL} and
\emph{RXTE} and of simultaneous fitting of all insturments on a given
satellite are given in Table~\ref{tab:mean_values}. All five
instruments' spectral parameters are in agreement at the 90\%
confidence level, except for the ISGRI determination of the power law
index. The ISGRI value of $\Gamma$ is significantly larger ($\Delta
\Gamma \sim$ 0.2$\pm$0.1). This discrepancy is $\sim0.1$ using OSA~5.0.

From the \emph{RXTE} observations, we have detected a 60\% increase in
the mean column depth to Cen~A between 2000 and 2003, and this
increase was not correlated to either the spectral index or the 2--10
keV flux. The increase in column depth was accompanied with a small
drop in iron line flux that is not significant at the 99\% confidence
level. By considering past satellite measurements of the absorbing
column, we note two episodes of $N_{\rm H}=1.0\times10^{23} \rm
cm^{-2}$ separated by $\sim$20 years, and speculate that variability
in the structure of the outer edge of the warped accretion disk could
explain the observed variability. Since the continuum shape and iron
line flux did not vary significantly, we suggest that they are
separate from each other and the intervening material.

Where, then, does the Fe K$\alpha$ line originate? Since the line
strength does not correlate with $N_{\rm H}$, we do consider it
unlikely that the Fe line is produced in the absorbing material.  The
similarity of the Cen A line parameters with those now seen with
\emph{Chandra} or \emph{XMM-Newton} in many AGN, low-luminosity
Seyferts, and high-luminosity QSOs, which all have narrow lines (to
the resolution of the observation) at an energy consistent with
emission by neutral Fe and equivalent widths of less than about
150\,eV \cite[see, e.g.,][]{Pounds02,Yaqoob04,Reynolds04,Jimenez05}
points at a similar origin for these features. The line parameters are
characteristic for a line origin as a fluorescence line in material
that is irradiated by X-rays. The possible location of the
line emitting region is either the outer regions of the
accretion disk or a medium separate from the disk, such as the torus
posited in unifying models for AGN. As shown by \citet{Ghisellini94}
and \citet{Leahy93}, for parameters typically assumed for the
torus and the central source, equivalent widths of 100\,eV are
expected. Alternatively, as discussed by \citet{Yaqoob01} and
\citet{Jimenez05}, such lines could also originate in the broad line
region, with the size of the region again being sufficiently large
that one would not expect a correlation between the flux from the
central source and the line strength.

We note, however, that the presence of a jet is one crucial difference
between radio quiet Seyfert galaxies and objects such as Cen A or
radio-loud QSOs. As has been recently shown for both Galactic black
hole candidates and for low luminosity AGN, a significant fraction of
the X-ray emission from these systems could also be explained by
synchrotron-self Comptonization radiation from the base of the radio
jet and thus not be due to thermal Comptonization \cite[][and
therein]{Markoff05,Falcke04}. The lack of a roll-over and reflection
component in Cen A, as opposed to Seyfert galaxies, could therefore
also be due to jet emission, with beaming or the small
footprint of the jet reducing the amount of reflection expected. In
these models, the advected flow at the base of the jet produces the
hot electrons and SSC radiation could extend to high energies. In
Seyfert galaxies, the jet could be less beamed/thermal, or thermal
Comptonization could be more dominant, resulting in the observed
roll-over. If this is the case, then the observed line emission could
also originate from a region that, due to beaming, is significantly
more illuminated by the jet than the disk. As we would not see the
illuminating radiation but only the Fe K$\alpha$ emission,
which is emitted isotropically, no correlation between the
continuum and the Fe line flux would be expected. In addition, such a
model could also explain the absence of hidden broad lines from radio
galaxies like Cen A, as the jet may sweep out the material that
otherwise would form the broad line region. It is beyond the
scope of this paper to quantify these effects, as they all strongly
depend on the unknown jet-kinematics at the base of the jet.

Naturally, the iron line emission could be a combination of these
effects, where the fraction transmitted through a torus (where the
line flux would be proportional to the primary flux) is small
compared to that reradiated at a distance. The lack of an iron edge
at 7.1 keV in the \textsl{XMM-Newton} and \textsl{Chandra} spectra
\citep{Evans04} as well as the \textsl{RXTE} data, further
complicates the question of the origin of the iron line in Cen~A.

\acknowledgments

RER acknowledges Chris Fragile for discussions on black hole disk
precession, and Phil Uttley on iron line and continuum variability.
RER acknowledges the support of NASA contract NAS5-30720 and NSF
international grants NSF\_INT-9815741 and NSF\_INT-0003773, as well as
the DAAD, for fostering the UCSD/T\"ubingen collaboration.

\clearpage

\appendix

\section{Background Subtraction}
\label{sec:bkgd}

The \emph{RXTE}/PCU2 was the only instrument of the 5 considered here
that required a small, but significant, correction to the estimated
background. The statistical uncertainty in the PCU2 3--60 keV source
detection is 0.12\%, whereas the SkyVLE background model is not
predicted to be accurate to better than a few percent in any given
observation.  The effect of this was evident in the initial spectral
fitting as a systematically negative flux above 20--30 keV. The PCU2
background is derived from a background model, which in turn is
derived from fitting background observations to mean particle counting
rates \citep{Jahoda04}. With the large collecting area of the PCU2
detector, statistical errors are quite small ($\sim0.1$\%) and thus
systematic uncertainties in the background model as applied to
specific observations can be seen in background-subtracted source
spectra.

The correction to the PCU2 background level was accomplished by an
iterative process of adjusting the amount of background and then
performing a $\chi^2$ fit to the basic spectral model ({\tt
phabs*(power + gauss(Fe))}) until the $\chi^2$ was minimized.
Technically, this was done through a series of FIT and RECOR commands
within XSPEC, with the estimated background data file also serving as
the correction file. Note, that this procedure is most effective when
the source flux spans the entire PCA energy range. Optimizing the
background estimate in this manner, however, may not be practical for
observations of steep spectrum sources. As seen in
Table~\ref{tab:pcu2_hexte_fits}, the corrections to the PCU2
background estimates (denoted by Cornorm) average about 3\% of
background. 

HEXTE background is measured nearly continuously during each
observation, and is a better estimate of the background than that from
a model. Over the 18--240 keV band, the uncertainty in the source flux
is $\sim0.5$\%. The mean HEXTE background corrections are 0.15\% with
1-$\sigma$ uncertainty of about 0.1\%. Thus, the HEXTE background
corrections are smaller than the detection significance and are
consistent with zero at 3-$\sigma$.

The \emph{INTEGRAL} instruments all derive their background estimates
from the background that is a natural part of coded mask image
deconvolution techniques. To use ISGRI as an example, the statistical
source detection uncertainty is $\sim0.15$\% of the source flux, which
is about the same as the uncertainty in the estimated background. None
of the \emph{INTEGRAL} spectra had the signature of a problem with
background subtraction (consistently positive or negative residuals
over a portion of the spectrum), and using a small fraction of the
estimated background as a correction to the subtracted background did
not yield any increase in the goodness of the fits. Thus, we do not
choose to alter the estimates of background of the individual
\emph{INTEGRAL} instruments.

\section{PCU2 Residuals}
\label{sec:pcu2_residuals}

By far the largest factor in finding an acceptable 3-60 keV fit to PCU2 data on
Cen~A was correcting the background estimate. In all cases, the
background was oversubtracted by 1.5--4.5\% (see
Appendix~\ref{sec:bkgd} above).

With the background subtraction optimized, significant residuals to
fitting the basic spectral model were seen at 8 keV (positive),
$\sim$30 keV (positive then negative), and near 60 keV (much smaller
positive then negative), as shown in
Fig.~\ref{fig:pcu2_residuals}(Top). The deviations from the model were
also reflected in the high $\chi^2_\nu$ values for Obs. 4-6. The 8 keV
residual can be attributed to imperfect modeling of the amount of
copper fluorescence from the Be/Cu collimator, as also indicated in
Cas~A calibration data (Appendix~\ref{sec:casa}).  The F-test yields
probabilities near a few percent for the need of the copper line in
the modeling.  The low significance of the line did not allow for
simultaneous determination of the flux and centroid energy, and thus
the line centroid energy was fixed at the copper K$_\alpha$ value of
8.04 keV. The flux of the residual copper line averaged about
$6\times10^{-5}$ photons cm$^{-2}$ s$^{-1}$.  This weak copper
K$_\alpha$ line was then included in subsequent analyses ({\tt
phabs*(power + gauss(Fe)) + gauss(Cu)}) to compensate for this effect.

This leaves the residuals near 30 and 60 keV. As mentioned by
\cite{Jahoda04} and seen in their Figure 20, the PCU2 background has
residual emission lines at 26, 30, and 60 keV from unflagged events
from the $^{241}$Am calibration source in each PCU. It is quite
possible that the PCU2 background model has an ever-so-slightly
different gain than the observed Cen~A data, and that this is the
origin of the residual seen in the Cen~A data at 60 keV. The fact that
the fitted intensities of a set of positive and negative gaussians around
60 keV are essentially equal, further supports this
hypothesis. 

The inclusion of a positive and negative set of Gaussian functions
offset by $\sim$2 keV at $\sim$59 and 61 keV ({\tt phabs*(power +
gauss(Fe)) + gauss(Cu) + gauss(Am+) + gauss(Am-)}), not only removed
the small residual at 60 keV, but it vastly improved the fit to the
data near the xenon K-edge at 33 keV
(Fig.~\ref{fig:pcu2_residuals}(Bottom)). This is due the fact that
inclusion of a positive or negative line near 60 keV produces K-escape
lines (positive or negative) near 30 keV.  Again, the low counting
rate of the lines ($\sim1\times10^{-3}$ photons cm$^{-2}$ s$^{-1}$)
and reduced PCU2 detector efficiency at 60 keV limited the number of
free line parameters in the fitting process.  After some
experimentation, it was found that having the positive line centroid
energy free, and the absolute value of the intensities of the two
lines set to be equal, allowed the fit to converge. The positive
Gaussian centroid averaged $\sim58.7$ keV and the negative Gaussian
centroid was fixed at 61 keV. We, therefore, will include an 8.04 keV
emission line and a fitted $\sim$58.7 keV emission line plus matching
negative Gaussian fixed at 61 keV to all fits to Cen~A.

In the case of Obs. 5, a significant negative residual is seen at
$\sim5$ keV. \citet{Jahoda04} discuss the calibration of the PCA with
respect to xenon L-edges and the resulting small residuals. Thus for
the Obs. 5, we included a narrow negative Gaussian in the model to
represent this systematic effect. The best-fit for the residual had a
centroid of 5.16$^{+0.14}_{-0.17}$ keV and flux of
$-1.4^{+0.4}_{-0.4}\times10^{-4}$ photons cm$^{-2}$ s$^{-1}$. The
other five observations were also fitted in this manner, but in all
five cases the flux of a fixed centroid at 5.16 keV was not
significant at the 90\% confidence level.
Table~\ref{tab:pcu2_hexte_fits} give the resulting best-fit spectral
parameters for the PCU2 observations of Cen~A. Finally, addition of an
overall systematic uncertainty of 0.3\% added to the data was required
to obtain $\chi^2_\nu\approx$1.

\section{PCU2 Iron Line Stability Using Cas A Observations}
\label{sec:casa}

In order to estimate the systematic uncertainties in Cen~A line
centroids, fluxes, and power law indices, we analyzed 11 Cas A
pointings taken over the history of \emph{RXTE} as a calibration
standard.  We used only PCU2 data and selected the energy range of
2.5--25 keV. The lower energy was chosen to be as low as possible
before encountering effects of the hardware low energy
threshold. Since the majority of the Cas~A observations had livetimes
of a few thousand seconds, no significant flux was seen above 25 keV,
and thus spectra were truncated at 25 keV. Again, systematics of 0.3\%
were added to the Cas~A data.

We modeled the Cas~A spectrum as a power law plus four Gaussians
representing emission from S, Ca, Ar, and Fe, and their respective
energies were fixed at 2.45, 3.12, and 3.87 keV for the first three
lines \citep{Holt1994}.  The Fe line centroid was free to assume its
best fit value, and its width was fixed at $\sigma$=0.01 keV. A fifth
line at 8.04 keV was added to represent incomplete modeling of the
copper collimator K$_\alpha$ fluorescence line
(Appendix~\ref{sec:pcu2_residuals}).  Our iterative procedure
described above also estimated that $\sim$5\%
corrections to the background were required. Table~\ref{tab:casa}
gives the best fit parameters for each of the 11 fits to the Cas~A
data, as well as the background correction value.

The observation dates span four of the five PCA gain/response epochs
--- none were available for Epoch 1. Epoch 5 did not entail a gain
change, but represents the time at which PCU0 lost its propane veto
layer. Thus, the response for PCU2 did not change from Epoch 4 to 5,
other than the slow gain variation with time for which PCARSP makes
correction. From this analysis we see that the low energy absorption
was constant as was the power law index. However, the flux from the
sulfur line at 2.45 keV (just below the analysis threshold of
$\sim$2.5 keV) abruptly doubled in going from Epoch 3 to 4, and
simultaneously, the Cas~A 2--10 keV flux rose from $\sim$(1.30 to
1.55)$\times10^{-10}$ ergs cm$^{-2}$ s$^{-1}$. The other noteable
change is in the Fe line centroid and to a lesser extent its line flux
in going from Epoch 2 to 3. The Fe line centroid changes by 1\% going
from 6.49 to 6.57 keV, while the flux drops from (6.2 to
6.0)$\times10^{-4}$ cm$^{-2}$ s$^{-1}$. Overall, the spectral
parameters are remarkably constant, with variations in the power law
index and Fe line centroid at $\sim$0.5\%, and in their fluxes at the
few percent level.

\section{JEM-X Residuals Study}\label{sec:jemx_lines}

As can be seen in Fig.~\ref{fig:jemx_spectra}, a broad structured feature
dominates the residuals from 7--9 keV in Obs. 5 \& 6. In order to try
to understand this feature, we added first one and then a
second narrow Gaussian component to the absorbed power law model with no
addition of systematic error (see Table~\ref{tab:jemx_lines}). For
Obs. 6, a third line at 19.3$\pm$1.0 keV was apparent in the
residuals, but was of minimal significance (F-test probability of
2$\times10^{-2}$). This line may be the molybdenum K-lines from the
collimator structure being fluoresced by charged particles or the
cosmic diffuse background. The ``line'' at 6.9--7.0 keV is at too high an
energy and is too strong to be the iron line seen by the PCA, and may
be due to imperfect modeling of the response where the background is
varying rapidly (see Fig. 7 of the JEM-X Analysis User Manual, Issue
4.2). The line at $\sim$8.6 keV is also seen in the background
spectrum and may be due to the copper edge at 8.98 keV.

We considered adding these lines to the model to address the effect of
systematics, as done for the PCU2. However, when the lines are
included in the fit, the column depth drops and the power law flattens
well beyond the \emph{RXTE} best-fit values. We thereupon choose to
abandon this method for dealing with systematics in JEM-X analysis and
added systematic errors to the spectral data to achieve
$\chi^2_\nu\approx1$. Also, the large residuals prevented any
meaningful estimates to be made of iron line fluxes from Cen~A.

\section{Spectral Accumulations for Sources with Large Angular
  Offsets\label{sec:150}}

An on-axis observation of NGC~4945 was made following the Cen~A
Obs. 5. This placed Cen~A $7\fdg32$ off-axis. The same procedures
were used to extract the spectrum of Cen~A in this case as were used
in the main analysis presented above. JEM-X, with its smaller field of
view, never viewed Cen~A at this time, and the SPI instrument only
detected Cen~A at 4$\sigma$. Consequently, only spectra from ISGRI
were accumulated. The ISGRI spectra were fit to a power law, and the
residuals revealed the same systematics as was the case when Cen~A was
viewed on-axis. The addition of 12\% systematic uncertainties were
necessary to obtain a $\chi^2_\nu$=1.09. The 20--100 keV flux ---
4.7$\times10^{-10}$ ergs cm$^{-2}$ s$^{-1}$ --- is completely
consistent with that seen in Obs. 5, and the best fit power law index,
$\Gamma$=1.79$\pm$0.16 is smaller, but consistent at the 90\%
confidence level, with that found in Obs. 5. From this we conclude
that no additional systematic effects are encountered in ISGRI
spectral accumulations and analysis of a source $\sim7^\circ$
off-axis.

\section{OSA~5.0 \emph{INTEGRAL} Results}
\label{sec:50}

The recent release of OSA 5.0 has made a marked improvement in
the JEM-X and ISGRI spectral results. The large residuals have been
supressed for the most part, and it is now unnecessary to add any
systematic errors to the Cen~A spectra from \emph{INTEGRAL}. ISGRI
spectral indices are now $\sim0.1$ larger than those from
\emph{RXTE}, and in the following spectral analysis results, the
ISGRI power law index was specified to be 0.1 larger than that for
JEM-X or SPI. In addition, the reduction of the large residual in
JEM-X near the iron line now allowed the addition of a narrow line
at 6.4 keV. Table~\ref{tab:50} gives the best fit spectral
parameters using OSA 5.0.


\clearpage

\clearpage

\clearpage

\clearpage

\clearpage

\clearpage

\begin{figure}[t]
\includegraphics[angle=270.,width=3.in]{f6a.ps}
\includegraphics[angle=270.,width=3.in]{f6b.ps}\\
\includegraphics[angle=270.,width=3.in]{f6c.ps}
\includegraphics[angle=270.,width=3.in]{f6d.ps}\\
\includegraphics[angle=270.,width=3.in]{f6e.ps}
\includegraphics[angle=270.,width=3.in]{f6f.ps}\\
\caption{Left: From top to bottom, PCU2 2.8--60 keV counts histograms for Obs. 4--6
  with instrumental features, background correction, and systematic
  errors included. The residuals to the best-fit histogram are
  displayed as $\chi$ versus energy below each histogram. Right:
  From top to bottom, the corresponding HEXTE 17--240 keV counts histograms for Obs. 4--6 with residuals
  displayed below each histogram. \label{fig:pcu2_hexte_last}}
\end{figure}

\clearpage

\begin{figure}[t]
\includegraphics[angle=90.,width=3.0in]{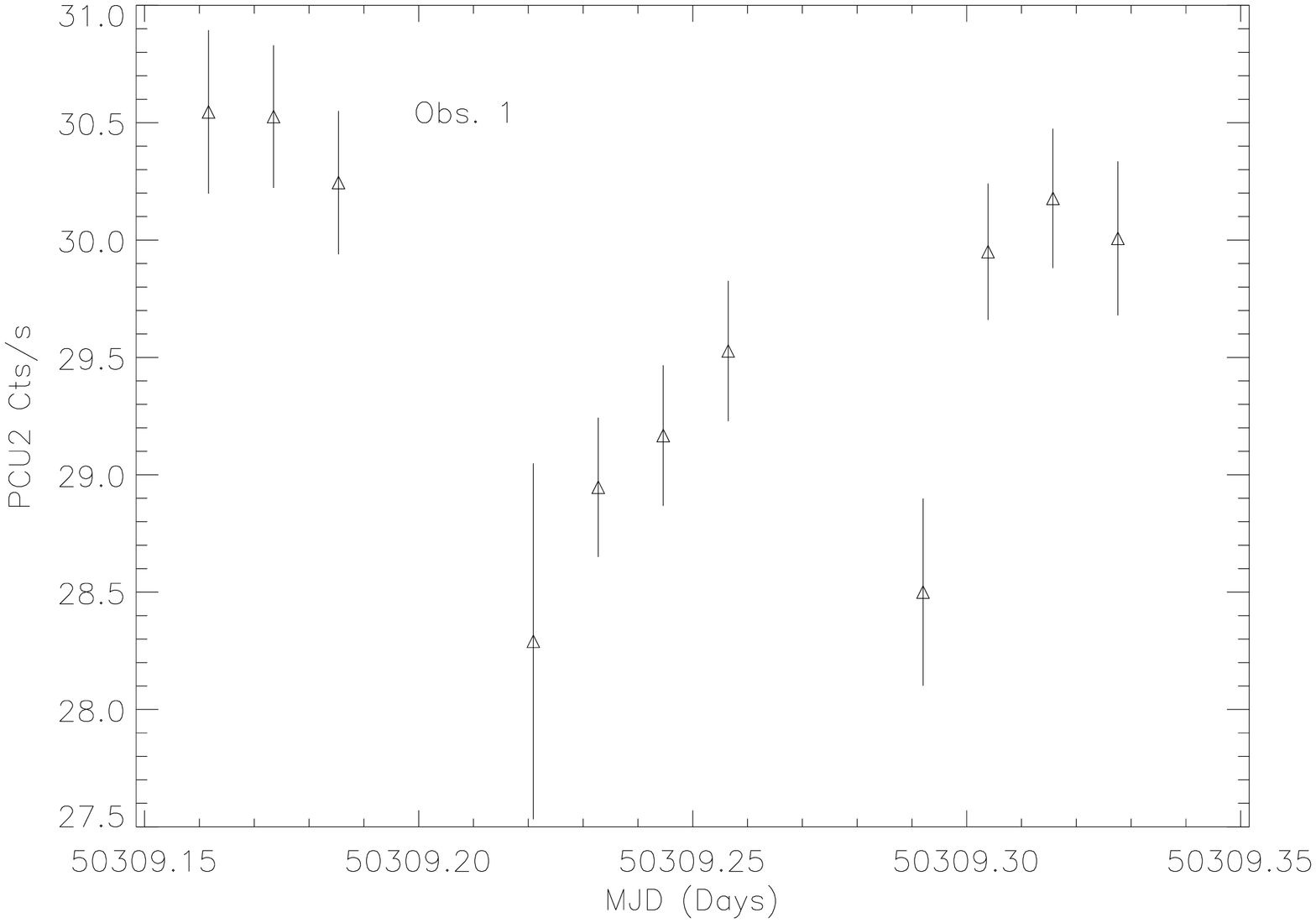}
\includegraphics[angle=90.,width=3.0in]{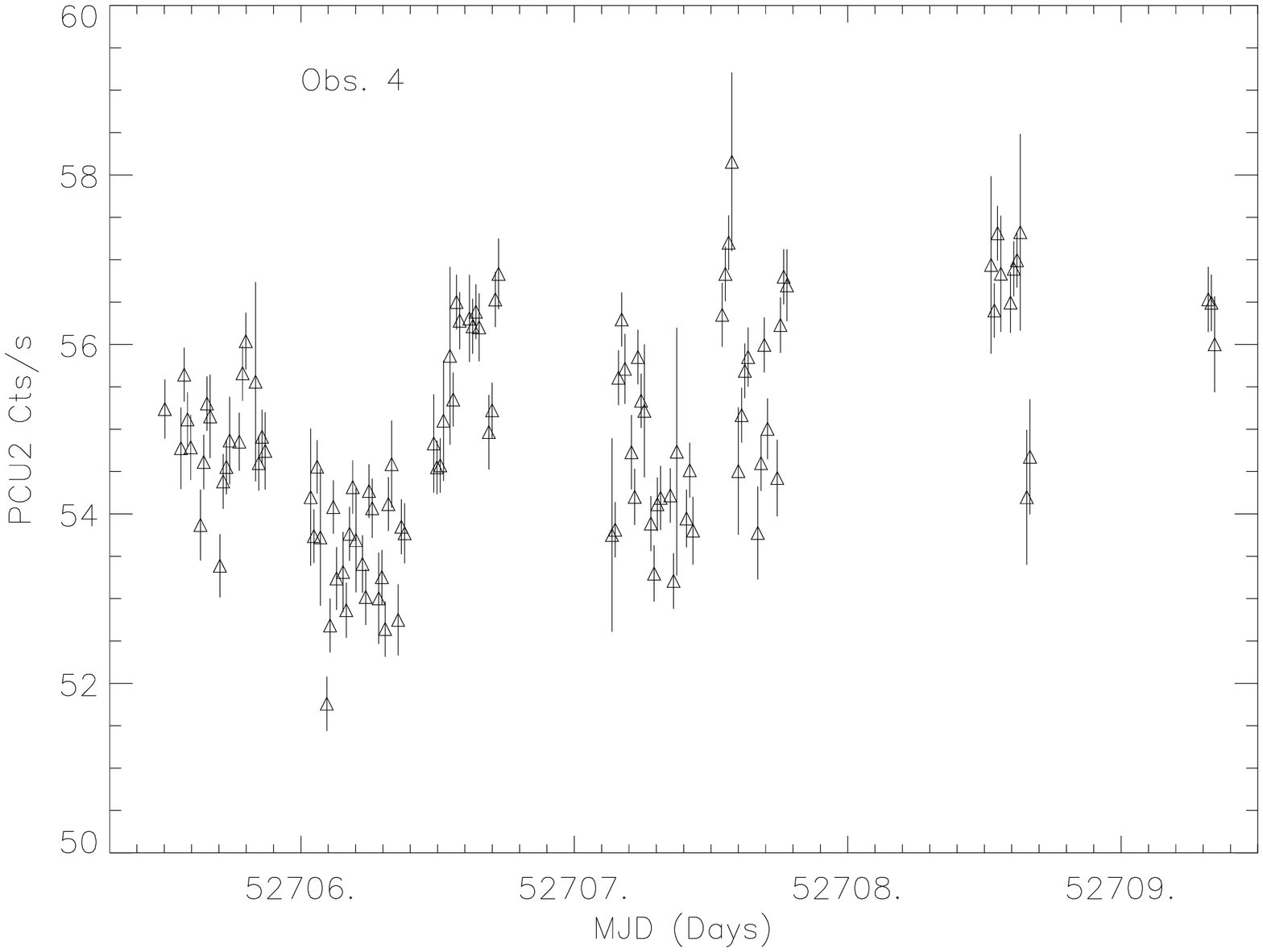}\\
\includegraphics[angle=90.,width=3.0in]{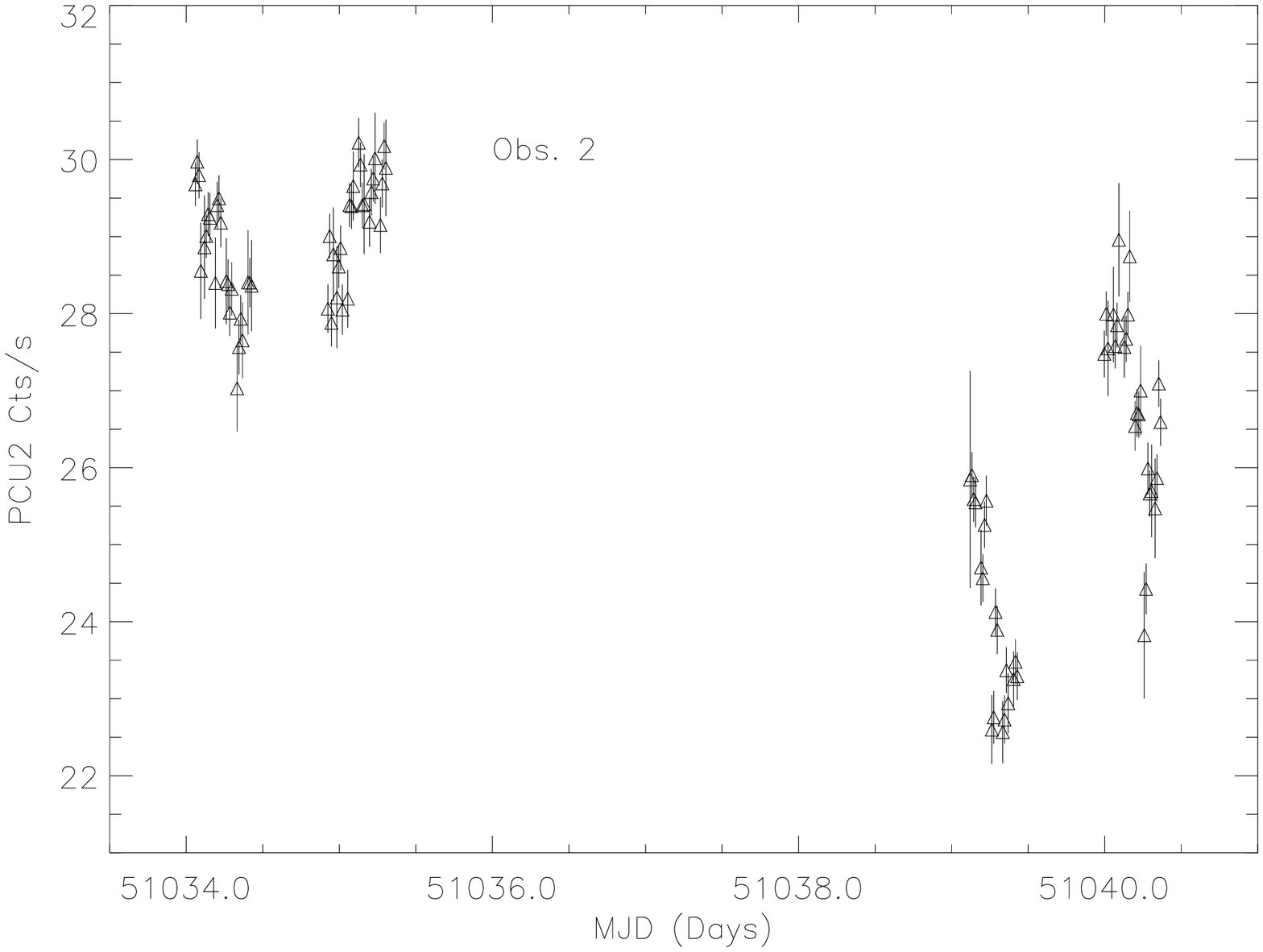}
\includegraphics[angle=90.,width=3.0in]{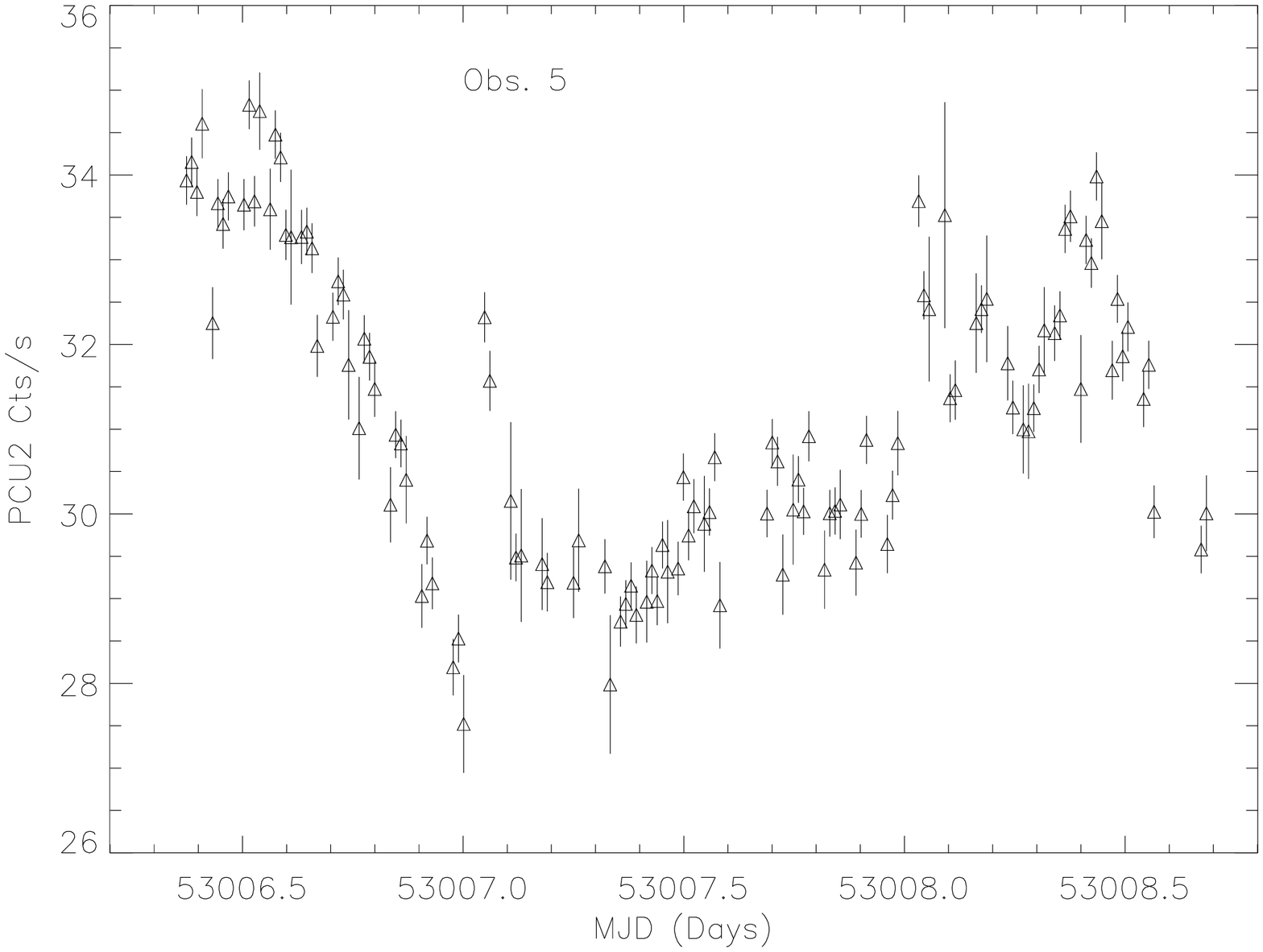}\\
\includegraphics[angle=90.,width=3.0in]{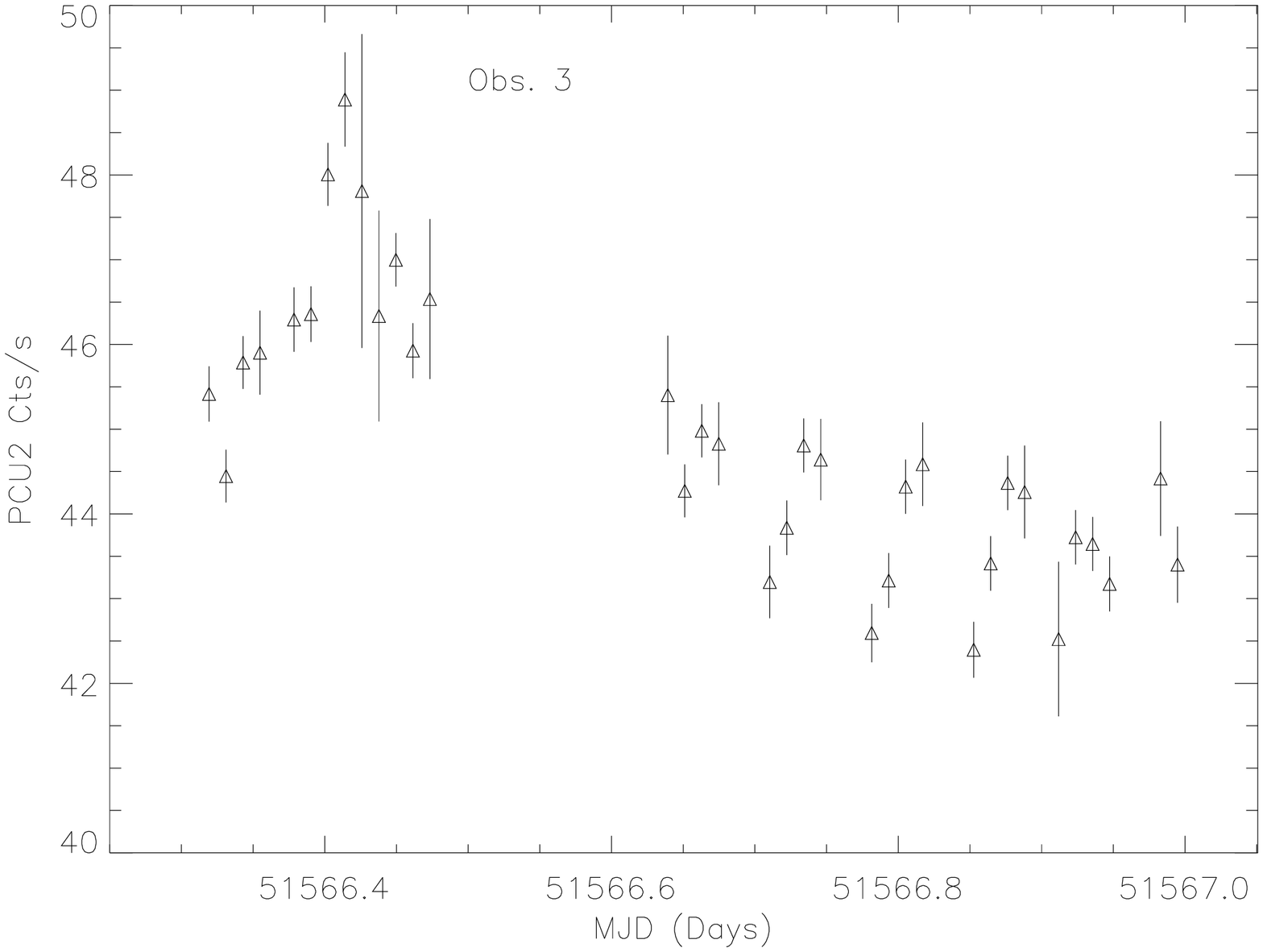}
\includegraphics[angle=90.,width=3.0in]{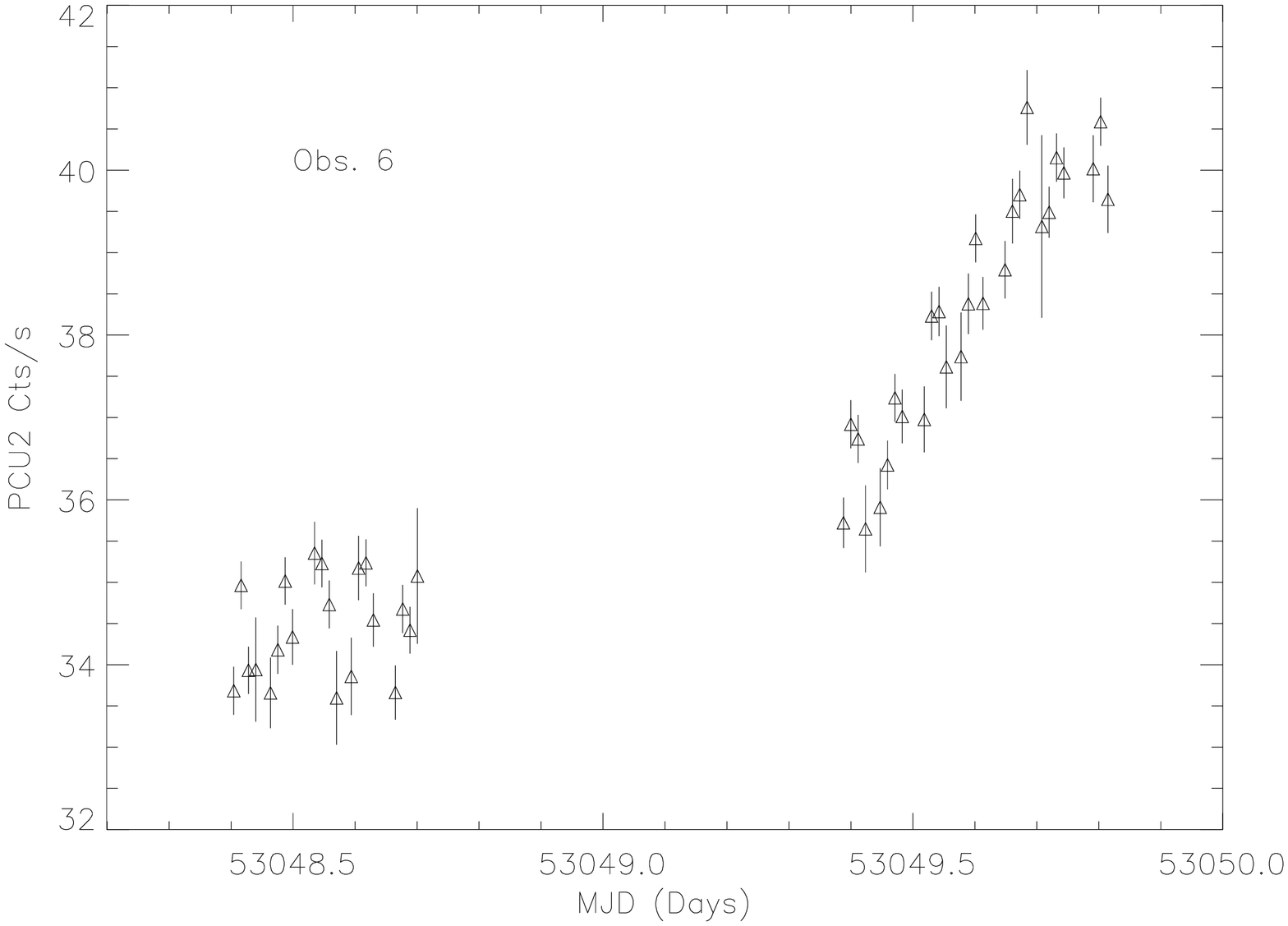}\\
\caption{The 2--60 keV PCU2 light curves for Cen~A for each
  observation. The rates are binned with 1024 s resolution, and have
  estimated background subtracted.\label{fig:lc}}
\end{figure}

\clearpage

\begin{figure}[t]
\centerline{\includegraphics[angle=90.,width=4.0in]{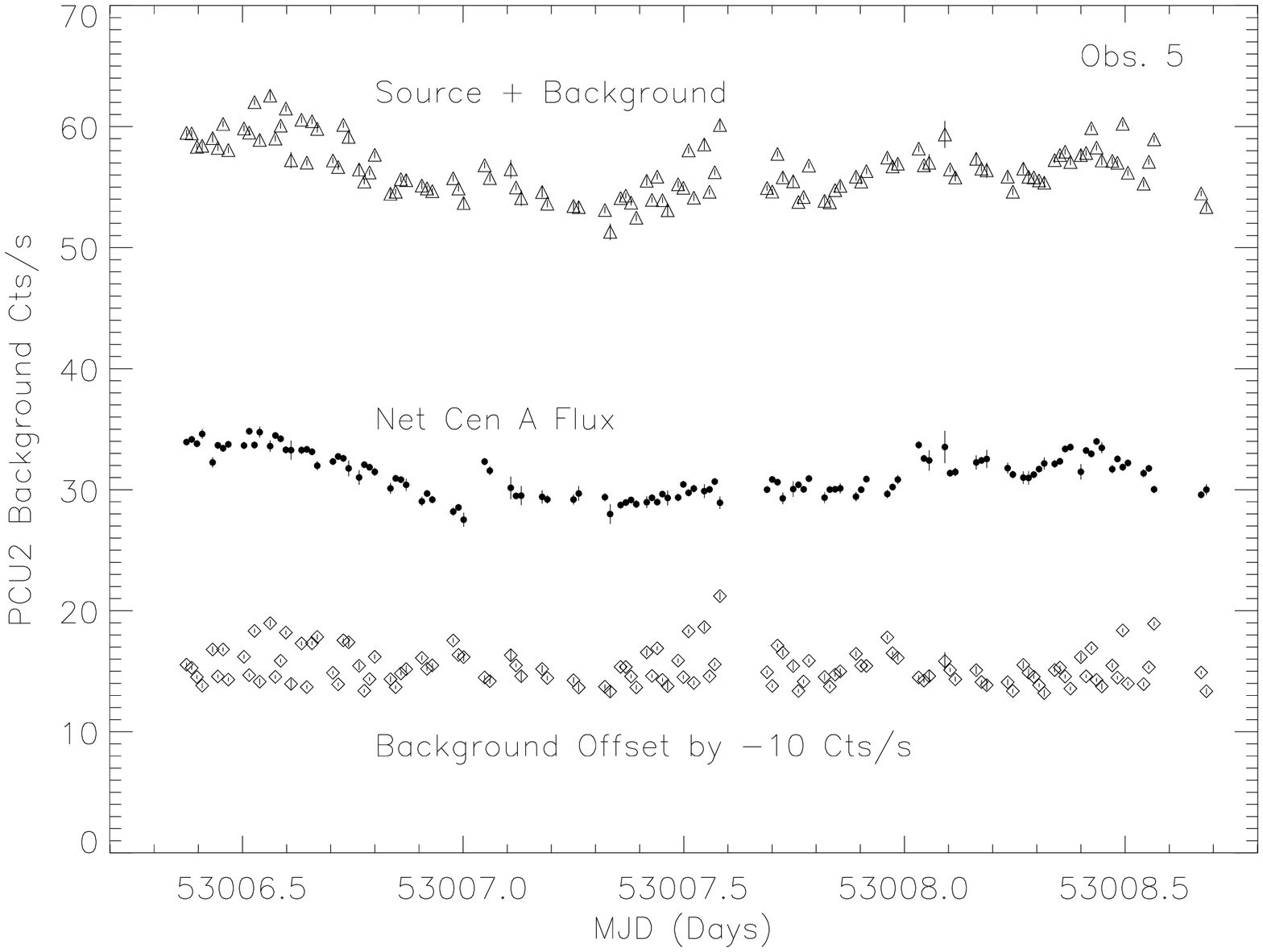}}
\includegraphics[angle=90.,width=3.0in]{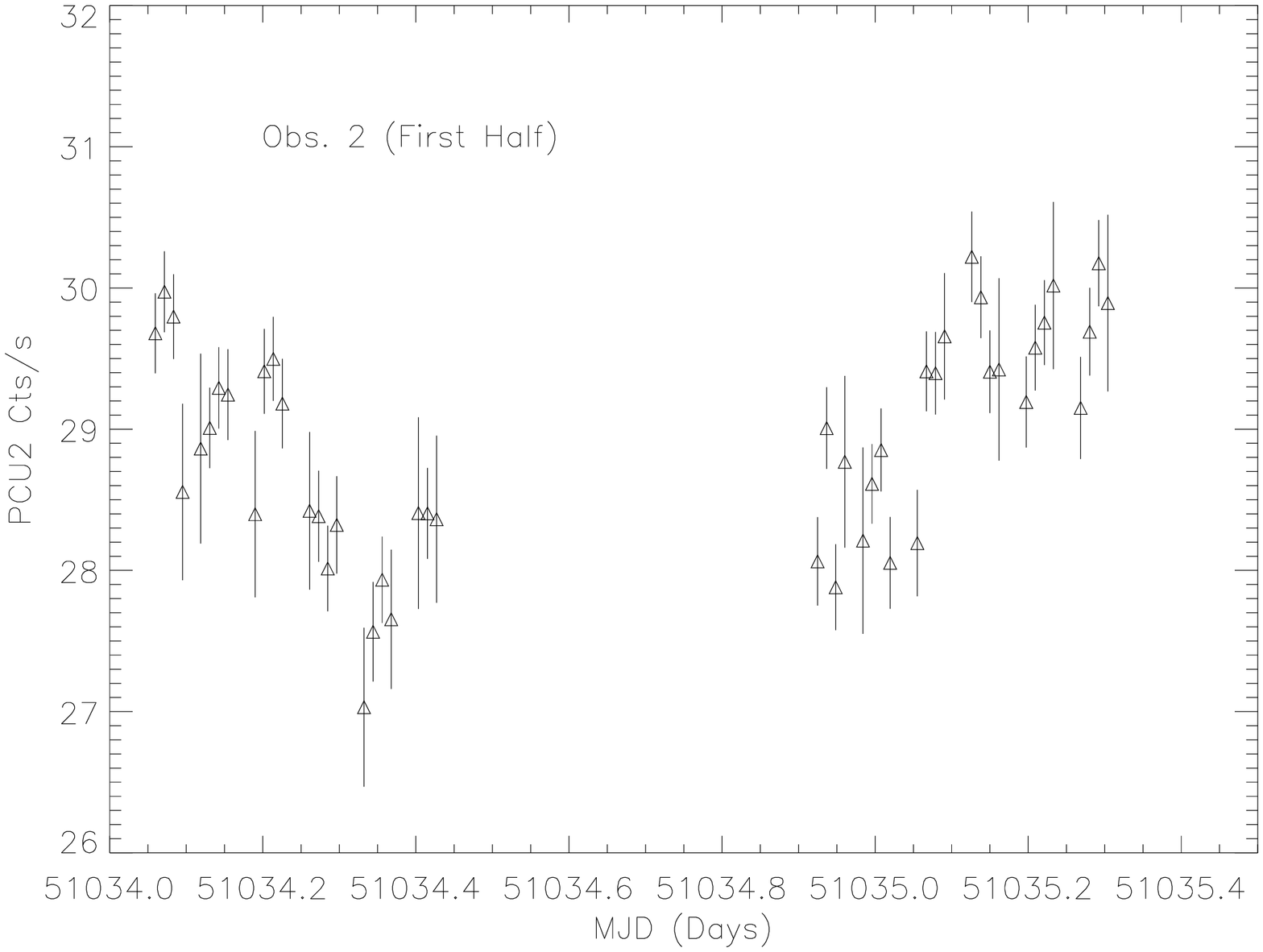}
\includegraphics[angle=90.,width=3.0in]{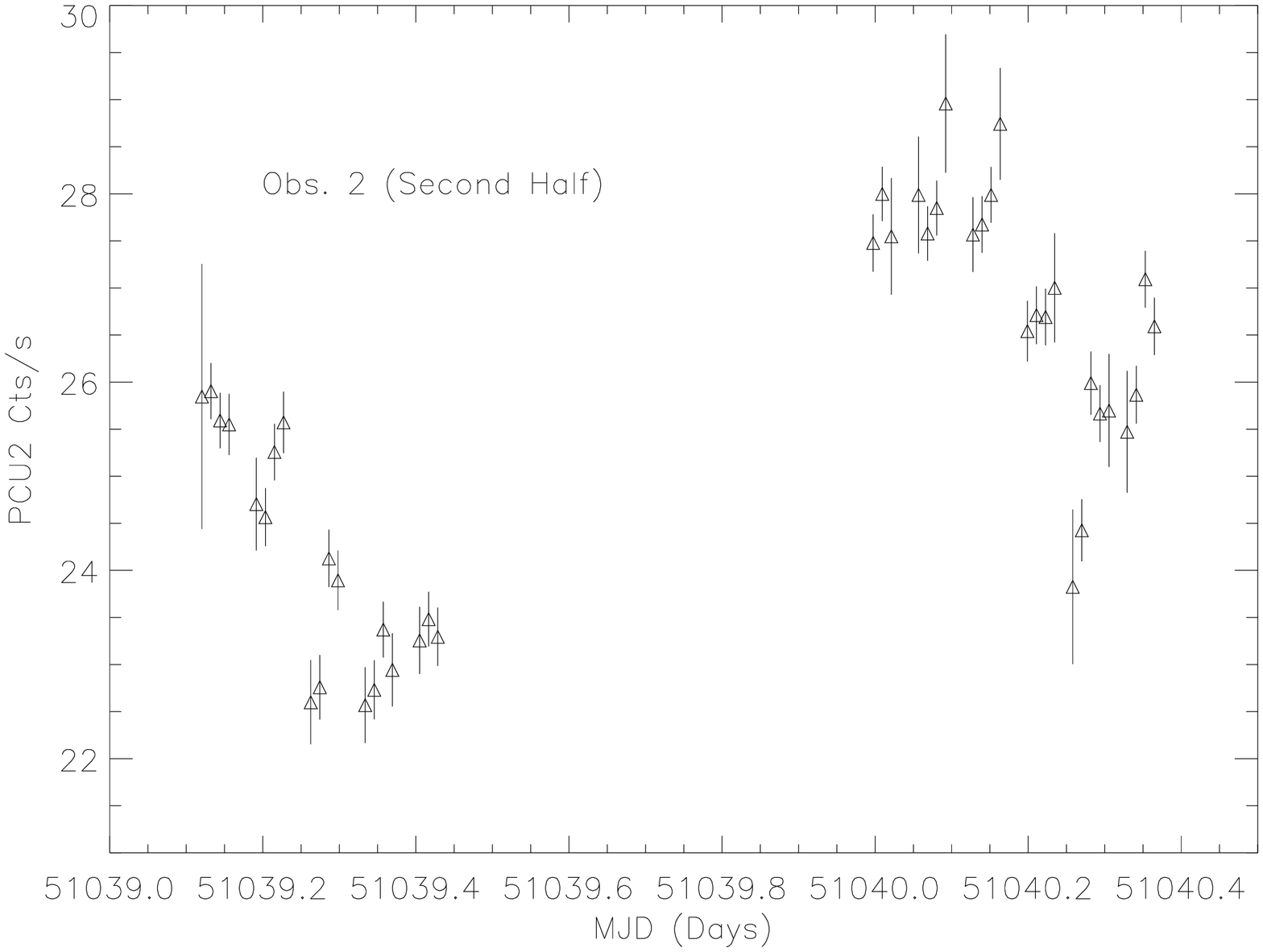}\\
\caption{Top: The 2-80 keV PCU2 light curves for Cen~A showing the
  accumulated detector rates as ``Source + Background'', from which the
  estimated background rates ``Background'' were subtracted, to yield
  the ``Net Cen~A Flux''. The background rates were offset by $-$10 c/s to
  provide separation from the net rates. Bottom: First day and a half
  and and last day and a half of Obs. 2. \label{fig:obs5lc}}
\end{figure}

\clearpage

\begin{figure}
\includegraphics[width=6.in]{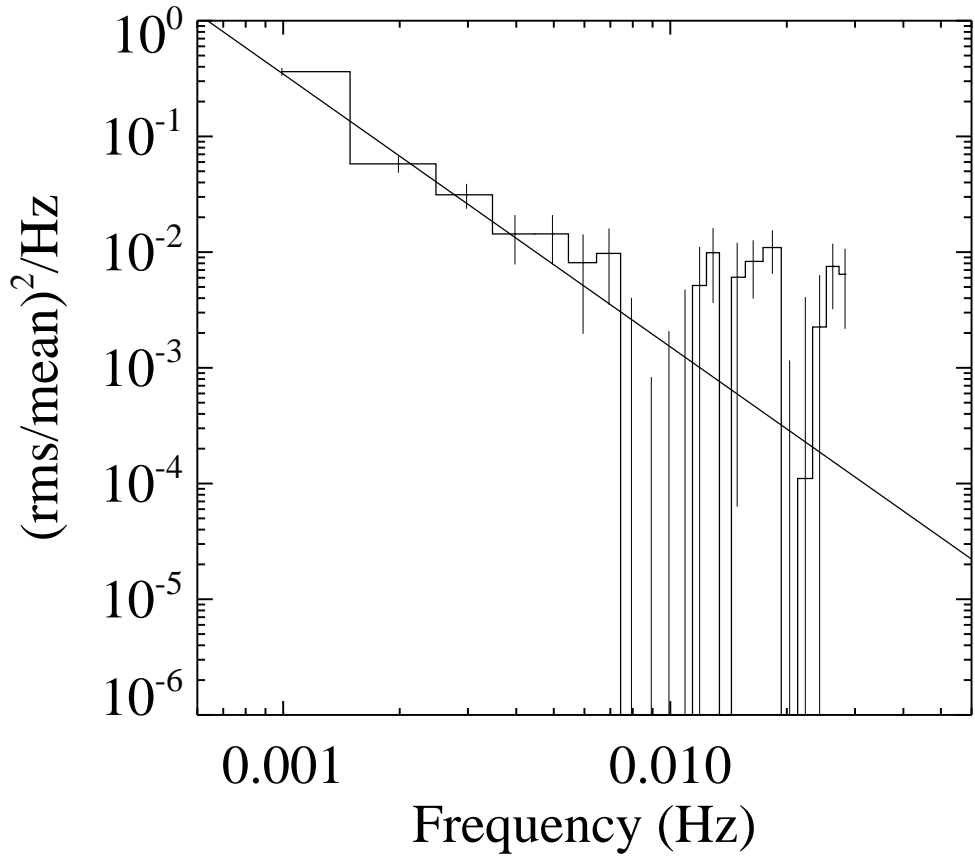}\\
\caption{The Cen~A RMS normalized power density function versus frequency.\label{fig:rms}}
\end{figure}

\clearpage

\begin{figure}[t]
  \includegraphics[angle=270.,width=6.in]{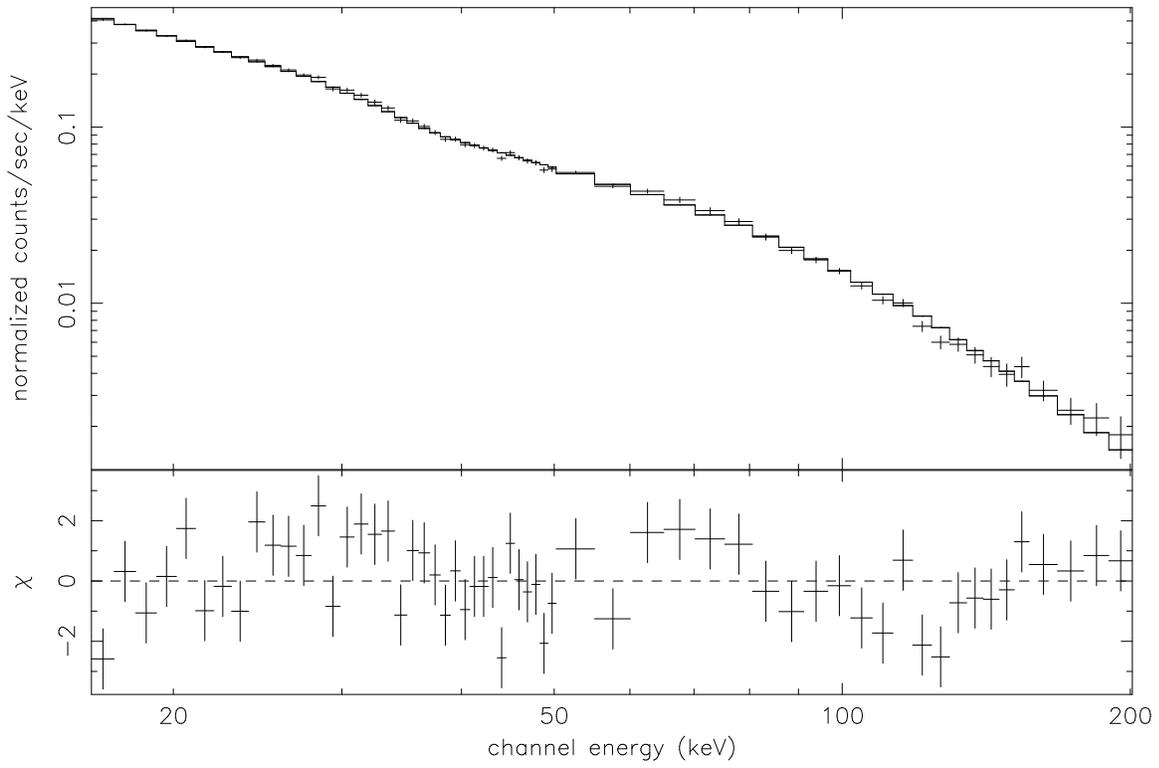}
  \caption{Power law fit to all 6 \emph{RXTE}/HEXTE observed summed
  together. Data above 50 keV are grouped by 5 channels until
  150 keV where the grouping increases to 10 channels.\label{fig:hexte_all6}}
\end{figure}

\clearpage

\begin{figure}[t]
\includegraphics[width=6.0in]{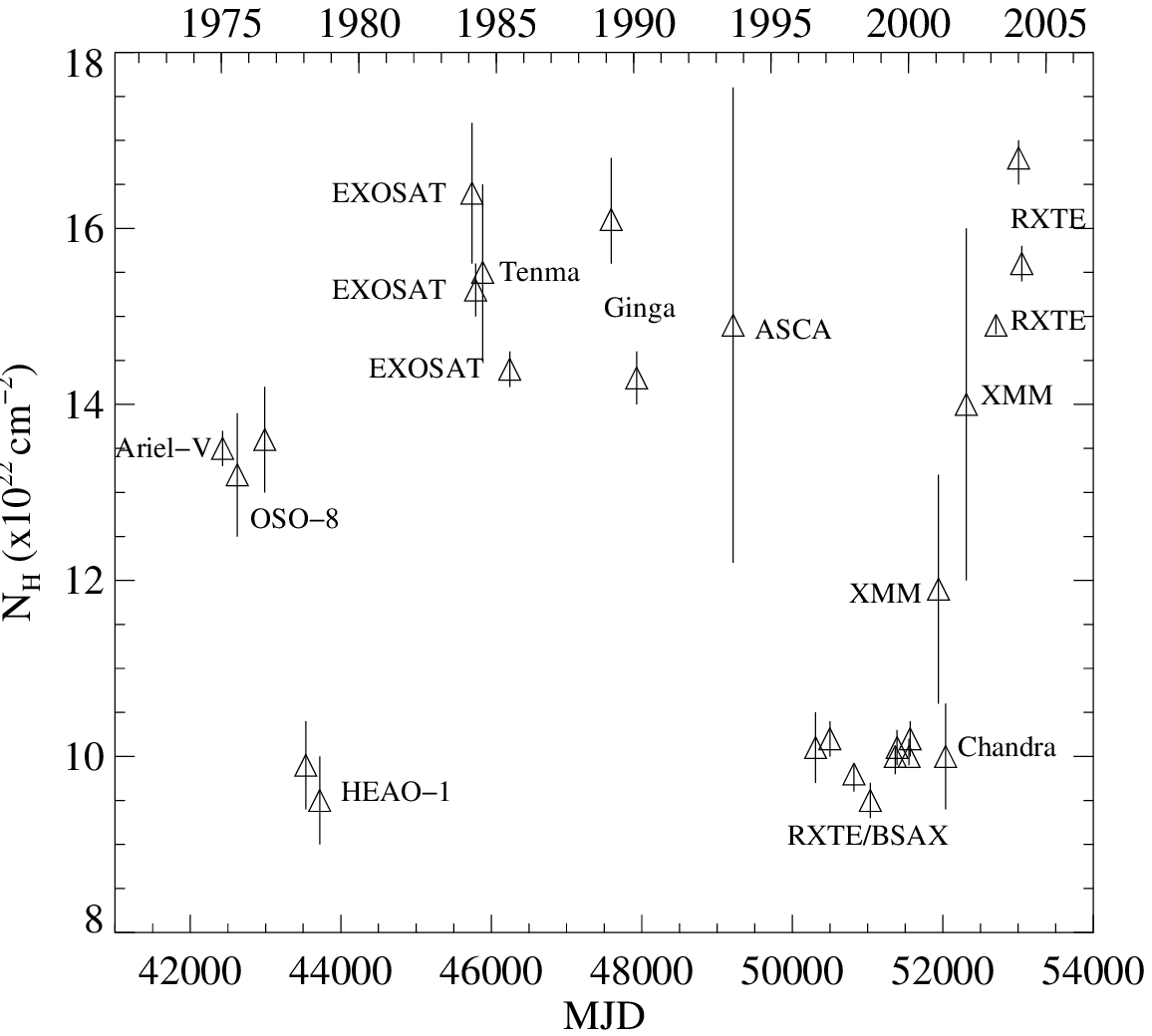}
\caption{History of line of sight column depth to Cen~A from 1975 to
  the present utilizing \emph{Ariel-V} \citep{Stark76}, \emph{OSO-8}
  \citep{Mushotzky78}, \emph{HEAO-1} \citep{Baity81}, \emph{EXOSAT}
  \citep{Morini89}, \emph{Tenma} \citep{Wang86}, \emph{Ginga}
  \citep{Miyazaki96}, \emph{ASCA} \citep{Sugizaki97}, \emph{BeppoSAX}
  \citep{Grandi03}, \emph{RXTE} (present paper),
  \emph{Chandra} \citep{Evans04}, and \emph{XMM-Newton} \citep{Evans04}. \label{fig:nh}}
\end{figure}

\clearpage

\begin{figure}[t]
\includegraphics[width=6.0in]{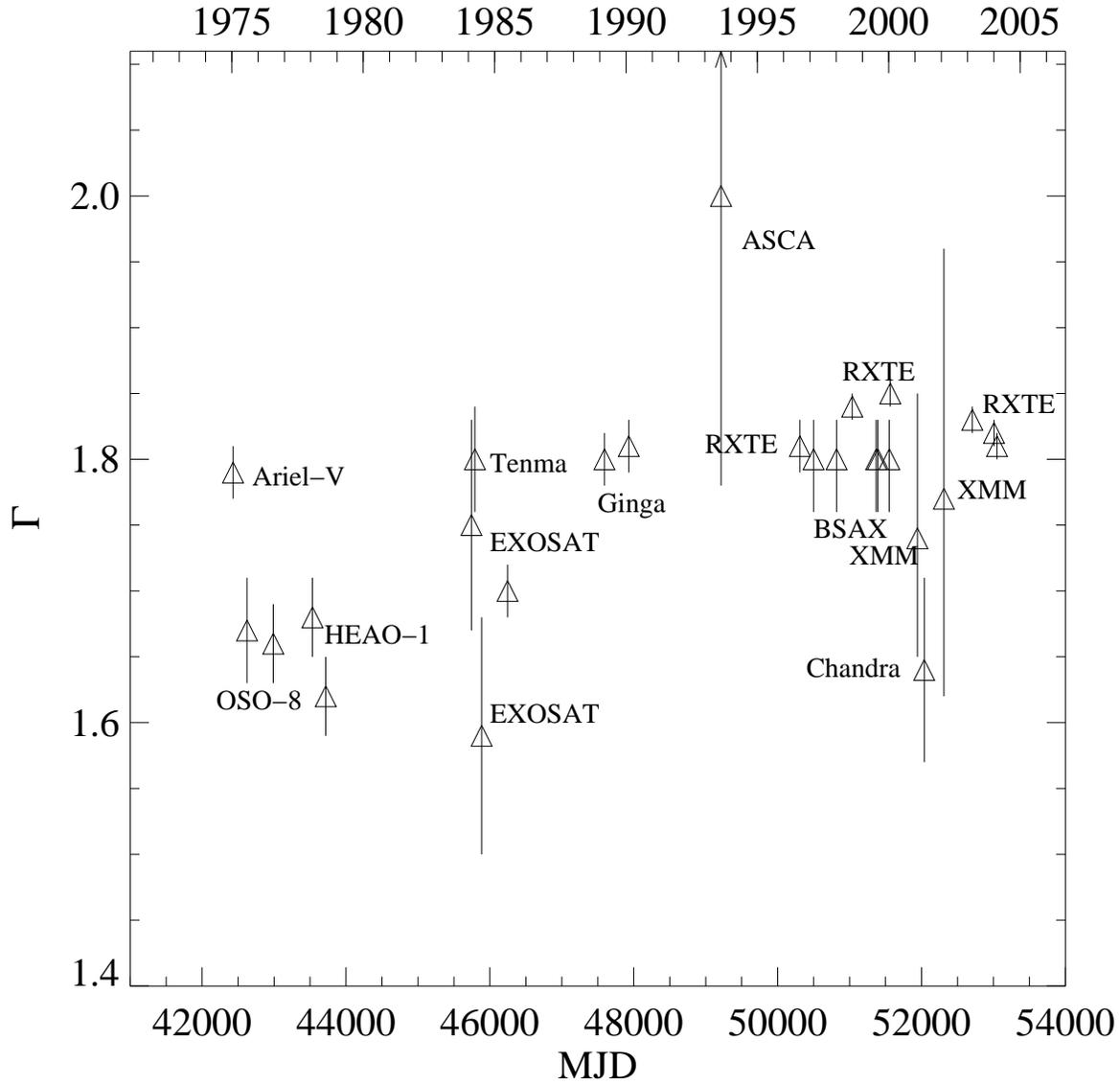}
\caption{The inferred power law index versus time for the past 30
  years. Missions/references are noted in Fig.~\ref{fig:nh}.\label{fig:index}}
\end{figure}

\clearpage

\begin{figure}[t]
\includegraphics[width=6.0in]{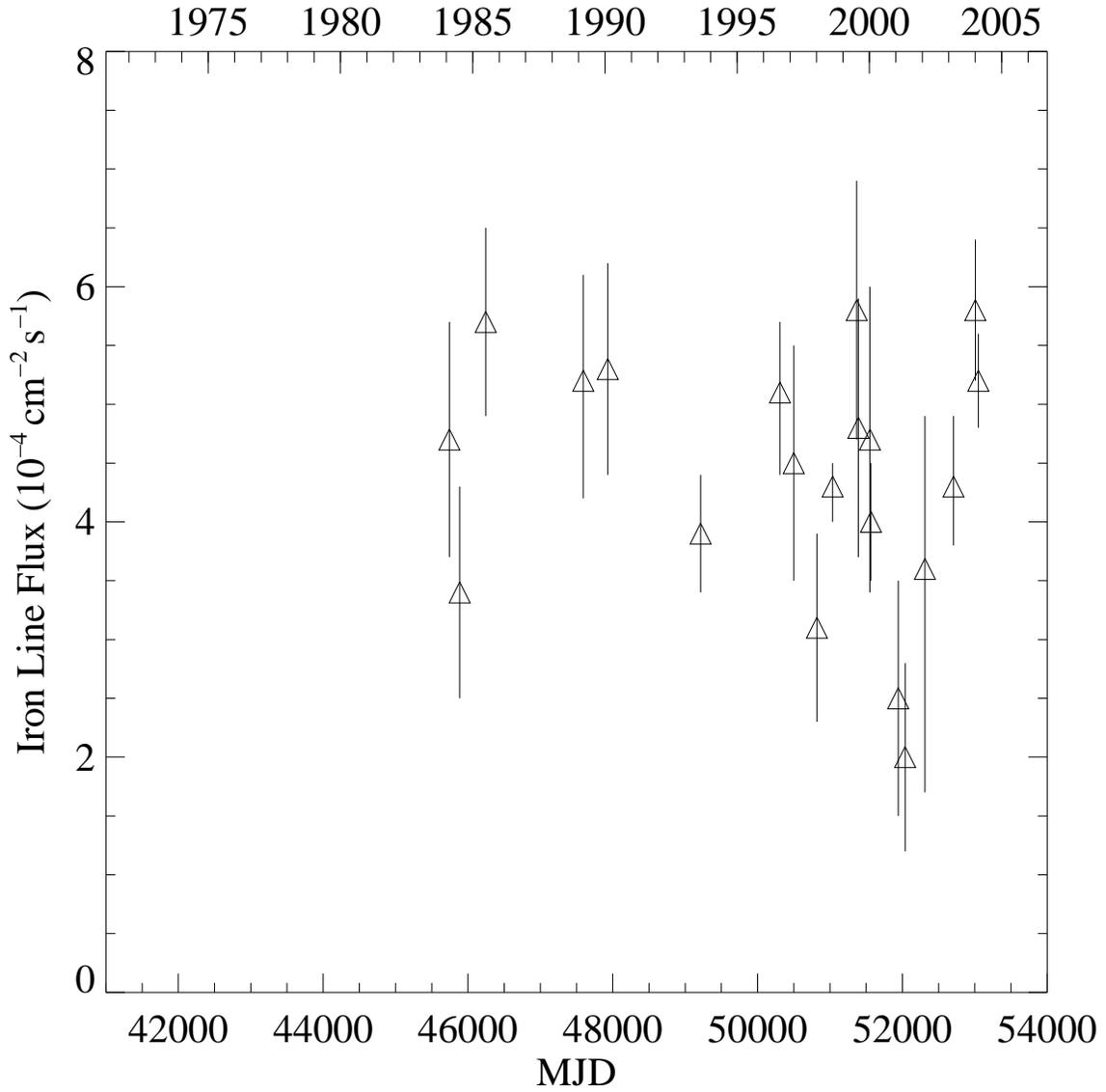}
\caption{Plot of iron line flux over
  the history of satellite measurements. The values are bracketed,
  in general between 4 and
  6$ \rm \times10^{-4}\; photons\; cm^{-2}\; s^{-1}$. Data points are
  from the same missions/references 
  as noted in Fig.~\ref{fig:nh}.\label{fig:index_flux}}
\end{figure}

\clearpage

\begin{figure}[t]
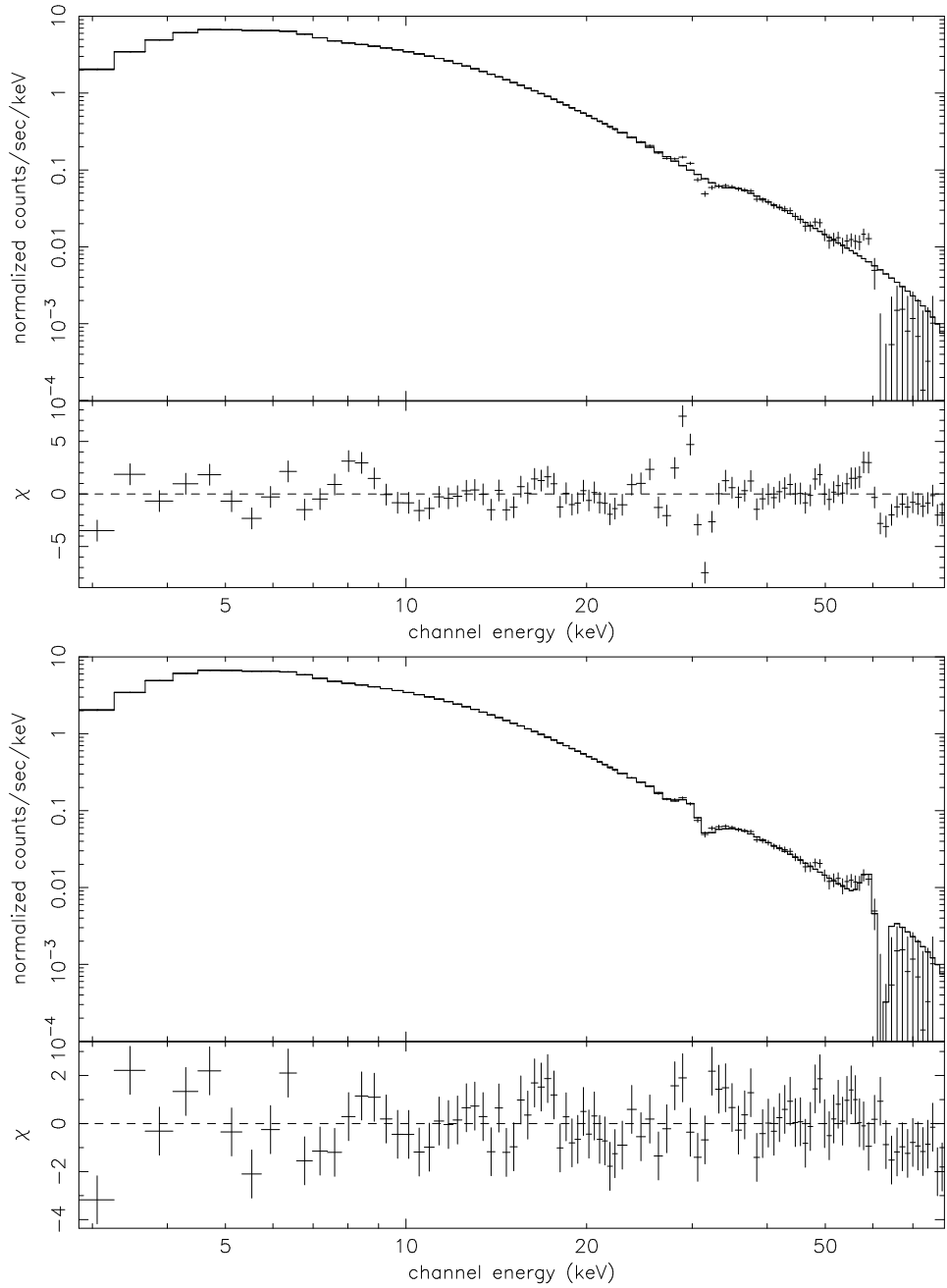

   \includegraphics[angle=270.,width=5.in]{f14a.ps}\\
   \includegraphics[angle=270.,width=5.in]{f14b.ps}
   \caption{Fit to PCU2 data from Cen~A, Observation 4, with just an absorbed power law and Fe line (Top), and in contrast,
   with an additional copper fluorescence line and a set of positive
   and negative Gaussians at 59 and 61 keV (Bottom). 
   Note the dramatic 
   effect on the residuals at the xenon K edge (34.6 keV).\label{fig:pcu2_residuals}}
\end{figure}


\clearpage

\begin{table}[t]
\footnotesize
\begin{center}
\caption{Details of the \emph{RXTE} and \emph{INTEGRAL} Monitoring Observations 
of Cen A\label{tab:obs}}
\begin{tabular}{ccccccc}
\tableline\tableline
\multicolumn{7}{l}{\underline{Instrument Exposures}}\\
Obs. Num.& Date & PCU2 & HEXTE & JEM-X & ISGRI & SPI \\ 
\tableline
1 & 1996 Aug. 14    & 10,528 &  6,785 & -- & -- & -- \\
2 & 1998 Aug. 9--15 & 67,872 & 44,688 & -- & -- & -- \\
3 & 2000 Jan. 23    & 25,088 & 16,132 & -- & -- & -- \\
4 & 2003 Mar. 7--11 & 86,624 & 58,574 & 78,005 & 101,030 & 92,942 \\
5 & 2004 Jan. 2--4  & 88,960 & 59,788 & 45,827 &  98,932 & 116,031 \\
6 & 2004 Feb. 13--14& 36,112 & 23,680 & 55,472 & 114,310 & 134,412 \\
\tableline
\multicolumn{7}{l}{\underline{Instrument Rates}}\\ 
& & PCU2 & HEXTE & JEM-X & ISGRI & SPI \\ 
Obs. Num.& MJD & 3-30 keV & 20-100 keV & 3-30 keV & 20-100 keV & 20-250 keV \\
\tableline
1 & 50309 & 29.32$\pm$0.08 & 3.84$\pm$0.18 & -- & -- & -- \\
2 & 51038 & 27.28$\pm$0.04 & 3.33$\pm$0.07 & -- & -- & -- \\
3 & 51566 & 44.06$\pm$0.06 & 5.84$\pm$0.10 & -- & -- & -- \\
4 & 52707 & 54.53$\pm$0.05 & 8.31$\pm$0.04 & 3.65$\pm$0.04 & 9.35$\pm$0.12 & (2.04$\pm$0.11)$\times$10$^{-2}$\\
5 & 53007 & 31.66$\pm$0.03 & 4.99$\pm$0.03 & 2.21$\pm$0.05 & 5.12$\pm$0.13 & (1.34$\pm$0.10)$\times$10$^{-2}$\\
6 & 53048 & 36.18$\pm$0.05 & 5.74$\pm$0.05 & 2.70$\pm$0.05 & 6.08$\pm$0.17 & (1.43$\pm$0.10)$\times$10$^{-2}$\\
\tableline
\tablecomments{All exposures are livetime in seconds; all rates are counts s$^{-1}$; 
the HEXTE rate is the sum of both clusters.}

\end{tabular}
\end{center}
\normalsize
\end{table}

\clearpage

\begin{deluxetable}{lccc}
\tabletypesize{\scriptsize}
\tablecaption{Best-fit JEM-X, ISGRI, \& SPI Spectral Parameters for Cen~A\label{tab:jemx_isgri_spi_fits}}
\tablewidth{0pt}
\tablehead{
\colhead{Parameter} & \colhead{Obs. 4} & \colhead{Obs. 5} &
\colhead{Obs. 6} 
}
\startdata
& JEM-X & (3--30 keV) &\\
\tableline
No Systematics &&&\\
$N_{\rm H}$\tablenotemark{a} & 15.4$^{+2.2}_{-2.1}$ & 14.5$^{+6.7}_{-5.5}$ &
14.3$^{+4.1}_{-3.6}$ \\
$\Gamma$ & 1.96$^{+0.08}_{-0.08}$ & 1.73$^{+0.20}_{-0.18}$ &
1.81$^{+0.13}_{-0.13}$ \\ 
Flux(2--10)\tablenotemark{b} & 2.96$^{+0.02}_{-0.06}$ & 1.52$^{+0.01}_{-0.13}$ &
1.94$^{+0.02}_{-0.11}$ \\
$\chi^2$/dof & 31.8/31=1.03 & 55.2/31=1.78 & 57.4/31=1.85 \\
\tableline
Systematics & 0.0\% & 25\% &17\% \\
$N_{\rm H}$\tablenotemark{a} & 15.4$^{+2.2}_{-2.1}$ & 12.5$^{+7.9}_{-6.4}$ &
13.4$^{+4.9}_{-4.2}$ \\
$\Gamma$ & 1.96$^{+0.08}_{-0.08}$ & 1.67$^{+0.24}_{-0.23}$ &
1.78$^{+0.16}_{-0.16}$ \\ 
Flux(2--10)\tablenotemark{b} & 2.96$^{+0.02}_{-0.07}$ & 1.51$^{+0.03}_{-0.19}$ &
1.93$^{+0.03}_{-0.16}$ \\
$\chi^2$/dof & 31.8/31=1.03 & 31.7/31=1.02 & 31.7/31=1.02 \\
\tableline
& ISGRI & (22.5--100 keV)& \\
\tableline
No Systematics\\
$\Gamma$ & 1.95$^{+0.04}_{-0.03}$ & 2.03$^{+0.06}_{-0.06}$ &
1.94$^{+0.05}_{-0.05}$ \\ 
Flux(20--100)\tablenotemark{c} & 9.04$^{+0.08}_{-0.09}$ & 4.95$^{+0.07}_{-0.14}$ &
5.88$^{+0.05}_{-0.11}$ \\
$\chi^2$/dof & 26.2/12=2.18 & 28.2/12=2.35 & 50.0/12=4.17\\
\tableline
Systematics & 3.5\% & 7.5\% & 9.5\% \\
$\Gamma$ & 1.97$^{+0.05}_{-0.05}$ & 2.07$^{+0.10}_{-0.10}$ &
1.99$^{+0.10}_{-0.11}$ \\ 
Flux(20--100)\tablenotemark{c} & 8.98$^{+0.06}_{-0.11}$ & 4.86$^{+0.07}_{-0.26}$ &
5.75$^{+0.09}_{-0.28}$ \\
$\chi^2$/dof & 13.3/12=1.10 & 11.5/12=0.96 & 11.2/12=0.94 \\
\tableline
& SPI & (20--250 keV)&\\
\tableline
No Systematics\\
$\Gamma$ & 1.79$^{+0.12}_{-0.11}$ & 1.66$^{+0.18}_{-0.19}$ &
1.89$^{+0.19}_{-0.19}$ \\ 
Flux(20--100)\tablenotemark{c} & 11.19$^{+0.20}_{-1.90}$ & 7.20$^{+0.21}_{-1.38}$ &
8.02$^{+0.15}_{-1.15}$ \\
$\chi^2$/dof & 13.3/18=0.74 & 8.0/18=0.45 & 19.8/18=1.10 \\

\enddata

\tablenotetext{a}{Absorption in units of 10$^{22}$ equivalent H atoms cm$^{-2}$}
\tablenotetext{b}{Flux in units of 10$^{-10}$ ergs cm$^{-2}$ s$^{-1}$ from 2--10 keV}
\tablenotetext{c}{Flux in units of 10$^{-10}$ ergs cm$^{-2}$ s$^{-1}$ from 20--100 keV}
\end{deluxetable}

\clearpage

\begin{deluxetable}{lcccccc}
\tabletypesize{\scriptsize}
\tablecaption{Best Fit PCU2 \& HEXTE Spectral Parameters for Cen~A\label{tab:pcu2_hexte_fits}}
\tablewidth{0pt}
\tablehead{
\colhead{Parameter} & \colhead{Obs 1} & \colhead{Obs 2} & \colhead{Obs 3} & \colhead{Obs 4} 
& \colhead{Obs 5} & \colhead{Obs 6}
}
\startdata
&&& PCU2 & (2.5--60 keV)&& \\
\tableline
$N_{\rm H}$\tablenotemark{a} & 10.41$^{+0.55}_{-0.50}$ & 9.69$^{+0.18}_{-0.22}$ & 10.26$^{+0.24}_{-0.26}$ 
& 15.03$^{+0.16}_{-0.15}$ & 17.17$^{+0.20}_{-0.20}$ & 15.81$^{+0.20}_{-0.25}$\\
$\Gamma$ & 1.829$^{+0.029}_{-0.032}$ & 1.854$^{+0.013}_{-0.009}$ &
1.857$^{+0.008}_{-0.016}$ & 1.838$^{+0.004}_{-0.008}$ & 1.841$^{+0.010}_{-0.011}$ 
& 1.830$^{+0.015}_{-0.007}$ \\
E(Fe)\tablenotemark{b} & 6.44$^{+0.11}_{-0.08}$ & 6.45$^{+0.04}_{-0.02}$ & 6.44$^{+0.04}_{-0.14}$ &
6.24$^{+0.06}_{-0.01}$ & 6.35$^{+0.05}_{-0.03}$ &
6.39$^{+0.05}_{-0.09}$ \\
Flux(Fe)\tablenotemark{c} & 4.6$^{+0.8}_{-0.8}$ & 4.0$^{+0.3}_{-0.4}$
& 3.7$^{+0.7}_{-0.7}$ & 4.0$^{+0.6}_{-0.6}$ & 5.2$^{+0.5}_{-0.5}$ &
4.8$^{+0.6}_{-0.7}$ \\
EW\tablenotemark{d} &  143 & 132 & 74
& 54 & 118 & 98 \\
Flux(2--10)\tablenotemark{e} & 1.76$^{+0.01}_{-0.01}$ & 1.69$^{+0.01}_{-0.01}$ & 2.74$^{+0.01}_{-0.01}$ 
& 3.23$^{+0.01}_{-0.01}$ & 1.84$^{+0.01}_{-0.01}$ & 2.13$^{+0.01}_{-0.01}$ \\
$\chi^2$ & 62/87=0.71 & 50/87=0.57 & 62/81=0.77 & 96/80=1.20 &
197/82=1.19 & 71/82=0.86\\
Cornorm\tablenotemark{f} & $-$1.55$\pm$0.32\% & $-$2.63$\pm$0.13\% & $-$2.99$\pm$0.22\% & $-$4.14$\pm$0.14\% 
& $-$4.99$\pm$0.16\% & $-$1.32$\pm$0.20\% \\
\tableline
&&& HEXTE & (17-240 keV)&& \\
\tableline
$\Gamma$ & 1.80$^{+0.11}_{-0.10}$ & 1.88$^{+0.05}_{-0.04}$ &
1.81$^{+0.04}_{-0.04}$ & 1.80$^{+0.01}_{-0.01}$ & 1.81$^{+0.02}_{-0.02}$ & 1.78$^{+0.02}_{-0.02}$ \\
Flux(20--100)\tablenotemark{g} & 4.33$^{+0.14}_{-0.32}$ & 3.90$^{+0.08}_{-0.11}$ & 6.90$^{+0.16}_{-0.14}$ 
& 9.89$^{+0.06}_{-0.06}$ & 5.91$^{+0.05}_{-0.06}$ & 6.88$^{+0.07}_{-0.05}$ \\
$\chi^2$ & 62/61=1.02 & 77/61=1.25 & 62/61=1.01 & 64/61=1.05 &
68/61=1.12 & 75/61=1.23\\
Cornorm\tablenotemark{f} & $-$0.06$\pm$0.15\% & $-$0.19$\pm$0.06\% & $-$0.36$\pm$0.11\% & $-$0.03$\pm$0.09\% 
& $-$0.12$\pm$0.09\% & $-$0.11$\pm$0.14\% \\
\enddata

\tablenotetext{a}{The PCU2 low energy absorption in units of 10$^{22}$ equivalent H atoms cm$^{-2}$}
\tablenotetext{b}{The iron line centroid and equivalent widths are in units of keV}
\tablenotetext{c}{The iron line flux in units of 10$^{-4}$ photons cm$^{-2}$ s$^{-1}$}
\tablenotetext{d}{The equivalent width is given in eV with respect to the absorbed spectrum}
\tablenotetext{e}{The PCU2 2--10 keV flux in units of 10$^{-10}$ ergs cm$^{-2}$ s$^{-1}$}
\tablenotetext{f}{The percentage additional background required}
\tablenotetext{g}{The HEXTE 20--100 keV flux in units of 10$^{-10}$ ergs cm$^{-2}$ s$^{-1}$}
\end{deluxetable}

\clearpage

\begin{deluxetable}{lcccccc}
\tabletypesize{\scriptsize}
\tablecaption{Best-fit Spectral Parameters for \emph{INTEGRAL} and
  \emph{RXTE} Observations of Cen~A\label{tab:integral_rxte_fits}}
\tablewidth{0pt}
\tablehead{
\colhead{Parameter} & \colhead{} & \colhead{Obs. 1} &
\colhead{} & \colhead{Obs. 2} & \colhead{} & \colhead{Obs. 3}   
}
\startdata
& & \emph{RXTE} & & \emph{RXTE} & & \emph{RXTE}  \\
N$_H$\tablenotemark{a} & & 10.4$^{+0.5}_{-0.5}$ & &
  9.7$^{+0.2}_{-0.2}$ & & 10.2$^{+0.2}_{-0.2}$ \\
$\Gamma$     & & 1.825$^{+0.027}_{-0.029}$ & &
  1.856$^{+0.014}_{-0.014}$ & & 1.853$^{+0.014}_{-0.014}$ \\
E(Fe)\tablenotemark{b} && 6.43$^{+0.16}_{-0.02}$ & & 6.41$^{+0.01}_{-0.01}$ & & 6.42$^{+0.05}_{-0.10}$\\
Flux(Fe)\tablenotemark{c} && 4.7$^{+0.8}_{-0.8}$ && 3.9$^{+0.4}_{-0.4}$ && 3.8$^{+0.6}_{-0.7}$\\
$\chi^2$/dof & & 123/149=0.83 & & 127/149=0.85 & & 127/144=0.89 \\
\tableline
Parameter & Obs. 4 & Obs. 4 & Obs. 5 & Obs. 5 & Obs. 6 & Obs.6 \\
\tableline
& \emph{INTEGRAL} & \emph{RXTE} & \emph{INTEGRAL} & \emph{RXTE} &  \emph{INTEGRAL} & \emph{RXTE} \\
N$_H$\tablenotemark{a} & 15.1$^{+1.5}_{-1.4}$ & 14.9$^{+0.1}_{-0.1}$ &  20.8$^{+5.1}_{-4.5}$ 
& 16.9$^{+0.2}_{-0.2}$ & 17.1$^{+3.4}_{-3.1}$ & 15.6$^{+0.2}_{-0.2}$ \\
$\Gamma$     & 1.949$^{+0.040}_{-0.040}$ & 1.830$^{+0.005}_{-0.006}$ &  1.955$^{+0.087}_{-0.085}$ & 
1.826$^{+0.012}_{-0.008}$ & 1.919$^{+0.081}_{-0.078}$ & 1.817$^{+0.012}_{-0.013}$ \\
E(Fe)\tablenotemark{b} && 6.24$^{+0.07}_{-0.01}$ & & 6.38$^{+0.02}_{-0.06}$ & & 6.36$^{+0.05}_{-0.03}$ \\
Flux(Fe)\tablenotemark{c} && 4.4$^{+0.5}_{-0.6}$ && 5.5$^{+0.5}_{-0.5}$ && 5.1$^{+0.9}_{-0.6}$\\
$\chi^2$/dof & 64/63=1.01 & 173/144=1.20 & 66/63=1.05 & 172/143=1.20 & 71/63=1.13 & 153/144=1.06 \\
\enddata
\tablenotetext{a}{The low energy absorption in units of 10$^{22}$ equivalent H atoms cm$^{-2}$}
\tablenotetext{b}{The iron line centroid in units of keV}
\tablenotetext{c}{The iron line flux in units of 10$^{-4}$ photons cm$^{-2}$ s$^{-1}$}

\end{deluxetable}

\clearpage

\begin{deluxetable}{lccccr}
\tabletypesize{\scriptsize}
\tablecaption{Tests for spectral steepening\label{tab:break}}
\tablewidth{0pt}
\tablehead{
\colhead{Obs. Num.} & \colhead{Model} & \colhead{$\Gamma_1$} & \colhead{$\rm E_C$\tablenotemark{a}} &
\colhead{$\Gamma_2$} & \colhead{$\chi^2$} 
}
\startdata
\multicolumn{2}{l}{PCU2\&HEXTE}\\
4 & power & 1.83$\pm$0.01 & --- & ---           & 173/144=1.20\\
  & bknpower & 1.83$\pm$0.01 & 150$^{+25}_{-40}$ & $\geq$1.90 & 164/142=1.15\\
  & cutoffpl & 1.82$\pm$0.01 & $\geq$1597 & --- & 182/143=1.27\\
&&&&&\\
5 & power & 1.83$\pm$0.01 & --- & --- & 172/143=1.20\\
  & bknpower & 1.83$\pm$0.01 & 94$^{+58}_{-30}$ & $\geq$1.93 & 164/141=1.16\\
  & cutoffpl & 1.82$\pm$0.02 & $\geq$677 & --- & 169/142=1.19\\
&&&&&\\
2 & power & 1.86$\pm$0.01 & --- & --- & 127/149=0.85\\
  & bknpower & 1.86$\pm$0.01 & 100 & 1.83$^{+0.96}_{-0.23}$ & 127/148=0.86\\
  & cutoffpl & 1.84$^{+0.02}_{-0.03}$ & $\geq$398 & --- & 126/148=0.85\\
\multicolumn{2}{l}{Summed HEXTE}\\
 & power & 1.77$\pm$0.01 & --- & --- & 83/58=1.43\\
 & bknpower & 1.76$\pm$0.01 & 70 & 1.85$\pm$0.06 & 78/57=1.37\\
 & cutoffpl & 1.72$\pm$0.03 & $\geq$668 & --- & 79/57=1.39\\
\enddata
\tablenotetext{a}{Break and Cutoff energies in keV}
\end{deluxetable}

\clearpage

\begin{deluxetable}{lcccccc}
\tabletypesize{\footnotesize}
\tablecaption{\emph{INTEGRAL} Observations of NGC 4945,
4507, and IC4329a\label{tab:ngc}} 
\tablewidth{0pt} 
\tablehead{
\colhead{Observation} & \colhead{Rate} & \colhead{Flux} &
\colhead{Rate} & \colhead{Flux} & \colhead{Rate} & \colhead{Flux}
}  
\startdata 
& \underline{NGC 4945} & \underline{$\Gamma$=1.88} & \underline{NGC 4507} & \underline{$\Gamma$=2.10} 
& \underline{IC4329a} & \underline{$\Gamma$=2.22}\\
&&&&&&\\ 
Obs. 4 & 1.20$\pm$0.07 & 1.71$^{+0.08}_{-0.07}$ & 1.60$\pm$0.08 & 2.31$^{+0.09}_{-0.08}$
& 1.67$\pm$0.16 & 2.69$^{+0.16}_{-0.27}$\\
Obs. 5 & 2.12$\pm$0.07 & 3.01$^{+0.12}_{-0.08}$ & 0.92$\pm$0.09 & 1.44$^{+0.10}_{-0.16}$ 
& 1.63$\pm$0.17 & 3.29$^{+0.19}_{-0.19}$\\ 
Obs. 6 & 1.76$\pm$0.06 & 2.52$^{+0.08}_{-0.10}$ & 0.62$\pm$0.09 & 1.16$^{+0.13}_{-0.14}$ 
& 1.13$\pm$0.21 & 3.10$^{+0.29}_{-0.27}$\\
\enddata 
\tablecomments{Rates are for 30--70 keV in c/s and Fluxes are
20--100 keV in units of $\rm 10^{-10} ergs\; cm^{-2}\; s^{-1}$. The
assumed power law index for calculating the flux is indicated above
the Flux columns.}
\end{deluxetable}

\clearpage

\begin{deluxetable}{lccccccc}
\tabletypesize{\footnotesize}
\tablecaption{Mean Spectral Parameters for Cen~A From \emph{INTEGRAL}
  and \emph{RXTE}\label{tab:mean_values}}
\tablewidth{0pt}
\tablehead{
\colhead{Parameter} & \colhead{PCU2} & \colhead{HEXTE} &
\colhead{JEM-X} & \colhead{ISGRI} & \colhead{SPI} & 
\colhead{\emph{RXTE}} & \colhead{\emph{INTEGRAL}} 
}
\startdata
$\langle N_{\rm H}\rangle$ & 15.9$^{+0.3}_{-0.2}$ & & 13.8$^{+5.5}_{-4.6}$ & & &
  15.8$\pm$0.2 & 17.7$^{+3.6}_{-3.3}$\\
$\langle\Gamma\rangle$  & 1.83$\pm$0.01 & 1.83$\pm$0.07 & 1.80$\pm$0.17 &
  2.01$\pm$0.09 & 1.78$\pm$0.17 & 1.82$\pm$0.01 & 1.94$\pm$0.07
\enddata
\tablecomments{$\langle N_{\rm H}\rangle$ in units of 10$^{22}$ cm$^{-2}$}
\end{deluxetable}

\clearpage

\begin{deluxetable}{lccccccc}
\tabletypesize{\scriptsize}
\rotate
\tablecaption{Best-fit PCU-2 Spectral Parameters for Cas A over the \emph{RXTE} Mission\label{tab:casa}}
\tablewidth{0pt}
\tablehead{
\colhead{Parameter} & \colhead{AO1} & \colhead{AO2} & \colhead{AO3} & \colhead{AO4-1} & \colhead{AO4-2} & \colhead{AO5} & \colhead{AO6}
}
\startdata
Date        & 4/1/96                 & 4/23/97                & 3/10/98                & 3/25/99                & 8/5/99                 & 9/6/00                 & 12/14/01\\
Epoch       & 2                      & 3                      & 3                      & 4                      & 4                      & 5                      & 5 \\
Index       & 3.28$^{+0.02}_{-0.02}$ & 3.29$^{+0.02}_{-0.03}$ & 3.30$^{+0.03}_{-0.03}$ & 3.29$^{+0.01}_{-0.02}$ & 3.29$^{+0.02}_{-0.03}$ & 3.32$^{+0.04}_{-0.04}$ & 3.29$^{+0.02}_{-0.03}$\\
Norm.       & 2.20$^{+0.09}_{-0.10}$ & 2.26$^{+0.07}_{-0.14}$ & 2.28$^{+0.14}_{-0.13}$ & 2.27$^{+0.06}_{-0.08}$ & 2.26$^{+0.08}_{-0.13}$ & 2.35$^{+0.20}_{-0.19}$ & 2.21$^{+0.08}_{-0.11}$\\
E(Fe)       & 6.492$^{+0.005}_{-0.003}$ & 6.572$^{+0.004}_{-0.003}$ & 6.575$^{+0.006}_{-0.003}$ & 6.568$^{+0.004}_{-0.004}$ & 6.573$^{+0.004}_{-0.004}$ & 6.562$^{+0.009}_{-0.066}$ & 6.568$^{+0.004}_{-0.006}$\\
Flux(Fe)    & 6.20$^{+0.13}_{-0.14}$    & 6.04$^{+0.12}_{-0.11}$  & 5.83$^{+0.17}_{-0.16}$  & 5.99$^{+0.12}_{-0.12}$    & 5.94$^{+0.14}_{-0.14}$    & 5.79$^{+0.25}_{-0.27}$    & 5.69$^{+0.14}_{-0.15}$\\
Flux(Cu)    & 4.8$^{+0.8}_{-0.8}$    & 5.7$^{+0.7}_{-0.7}$    & 5.5$^{+1.0}_{-1.0}$    & 5.3$^{+0.7}_{-0.7}$    & 5.2$^{+0.9}_{-0.4}$    & 4.4$^{+1.5}_{-1.6}$    & 5.6$^{+0.9}_{-0.9}$\\
Flux(S)     & 5.7$^{+0.4}_{-0.4}$    & 6.2$^{+1.5}_{-1.3}$    & 4.3$^{+1.5}_{-1.5}$    & 12.1$^{+0.4}_{-0.4}$    & 11.9$^{+0.5}_{-0.4}$    & 11.4$^{+0.7}_{-0.7}$    & 12.4$^{+0.4}_{-0.4}$\\
Flux(Ca)    & 2.2$^{+0.7}_{-0.7}$    & 1.5$^{+0.7}_{-0.4}$    & 1.7$^{+0.7}_{-0.6}$    & 1.7$^{+0.5}_{-0.4}$    & 2.0$^{+0.6}_{-0.6}$    & 1.1$^{+0.9}_{-0.9}$    & 2.2$^{+0.6}_{-0.5}$ \\
Flux(Ar)    & 5.1$^{+0.9}_{-0.9}$    & 2.8$^{+1.5}_{-1.1}$    & 3.6$^{+1.7}_{-1.6}$    & $\leq$0.5              & $\leq$1.33             & $\leq$2.3              & $\leq$0.9 \\
$\chi^2$    & 38.4/51=0.75           & 43.2/45=0.96           & 24.2/45=0.54           & 49.6/42=1.18           & 35.5/42=0.84           & 34.0/42=0.81           & 44.8/42=1.07\\
Flux(2-10)  & 1.32                   & 1.35                   & 1.27                   & 1.56                   & 1.40                   & 1.53                   & 1.55\\
Bkgd. Corr. & -5.5$\pm$0.9\%         & -5.3$\pm$0.8\%         & +0.0$\pm$1.1\%         & -3.6$\pm$0.9\%         & -5.0$\pm$0.9\%         & +4.6$\pm$2.1\%         & -2.8$\pm$1.2\%\\
Livetime    & 3616                   & 4992                   & 2608                   & 4624                   & 3312                   & 944                    & 2784\\
\tableline
\tableline
Parameter & AO8-1   & AO8-2  & AO9-1   & AO9-2 & Mean     & RMS    & RMS/Mean\\
\tableline
Date        & 5/16/03                & 5/22/03                & 6/26/04                & 7/1/04    \\
Epoch       & 5                      & 5                      & 5                      & 5         \\
Index       & 3.29$^{+0.03}_{-0.03}$ & 3.27$^{+0.03}_{-0.03}$ & 3.30$^{+0.02}_{-0.03}$ & 3.27$^{+0.03}_{-0.03}$              & 3.290    & 0.014  & 0.4\% \\
Norm.       & 2.18$^{+0.06}_{-0.12}$ & 2.09$^{+0.11}_{-0.11}$ & 2.22$^{+0.10}_{-0.12}$ & 2.06$^{+0.12}_{-0.12}$              & 2.22     & 0.08   & 3.8\% \\
E(Fe)       & 6.563$^{+0.006}_{-0.061}$ & 6.560$^{+0.007}_{-0.059}$ & 6.500$^{+0.061}_{-0.007}$ & 6.515$^{+0.050}_{-0.019}$  & 6.550    & 0.031  & 0.5\% \\
Flux(Fe)    & 5.81$^{+0.16}_{-0.15}$    & 5.93$^{+0.14}_{-0.15}$    & 5.71$^{+0.16}_{-0.16}$    & 5.83$^{+0.16}_{-0.17}$     & 5.89     & 0.15   & 2.6\% \\
Flux(Cu)    & 5.5$^{+1.0}_{-1.0}$    & 4.9$^{+0.9}_{-0.9}$    & 4.7$^{+0.9}_{-0.9}$    & 5.0$^{+1.0}_{-1.0}$                 &      & \\
Flux(S)     & 12.5$^{+0.5}_{-0.5}$   & 12.9$^{+0.4}_{-0.4}$   & 12.3$^{+0.4}_{-0.4}$   & 12.9$^{+0.5}_{-0.5}$                &          &  \\
Flux(Ca)    & 2.2$^{+0.6}_{-0.6}$    & 2.7$^{+0.6}_{-0.6}$    & 1.7$^{+0.6}_{-0.5}$    & 2.9$^{+0.6}_{-0.6}$                 &          &  \\
Flux(Ar)    & $\leq$1.6              & 1.3$^{+1.0}_{-1.0}$    & $\leq$1.2              & 1.6$^{+0.8}_{-1.1}$                 &          &  \\
$\chi^2$    & 39.8/42=0.95           & 53.7/42=1.28           & 43.2/42=1.03           & 57.3/42=1.36 \\
Flux(2-10)  & 1.55                   & 1.55                   & 1.54                   & 1.54                                &          &  \\
Bkgd. Corr. & -1.6$\pm$1.3\%         & -4.1$\pm$1.1\%         & -3.6$\pm$1.2\%         & -3.7$\pm$1.2\% \\
Livetime    & 2496                   & 3008                   & 2752                   & 2400 \\
\enddata

\tablecomments{All errors are 90\%; E(Fe) in units of keV and $\sigma$(Fe)=0.01 keV; E(Cu)=8.04 keV, E(S)=2.45 keV, E(Ca)=3.87 keV, and E(Ar)=3.12 keV; 
Flux(S) in units of 10$^{-2}$ cm$^{-2}$ s$^{-1}$; Flux(Fe, Ca, Ar) in units of 10$^{-3}$ cm$^{-2}$ s$^{-1}$; Flux(Cu) in units of 10$^{-4}$ cm$^{-2}$ s$^{-1}$; 
Flux(2-10) in units of 10$^{-9}$ ergs cm$^{-2}$ s$^{-1}$; Livetime in s; and fitted energy range was 3.5--25 keV.
}

\end{deluxetable}

\clearpage

\begin{deluxetable}{lcccc}
\tabletypesize{\scriptsize}
\tablecaption{Best-fit Spectral Parameters for JEM-X Residuals Study
  \label{tab:jemx_lines}}
\tablewidth{0pt}
\tablehead{
\colhead{Parameter} & \colhead{No Lines} & \colhead{One Line} &
\colhead{Two Lines} &  \colhead{Systematics}    
}
\startdata
&& Obs. 4 && 0\%\\
\tableline
$N_{\rm H}$ & 15.4$^{+2.2}_{-2.1}$   & & &  15.4$^{+2.2}_{-2.1}$\\
$\Gamma$  & 1.96$^{+0.08}_{-0.08}$ & & &  1.96$^{+0.08}_{-0.08}$\\
Flux(2-10)& 2.95$^{+0.02}_{-0.07}$ & & &  2.96$^{+0.02}_{-0.06}$\\
$\chi^2$  & 31.8/31=1.03           & & &  31.8/31=1.03\\
\tableline
&& Obs. 5 && 25\%\\
\tableline
$N_{\rm H}$ & 14.5$^{+6.7}_{-5.5}$ & 8.5$^{+6.3}_{-5.2}$ & 6.1$^{+6.0}_{-5.6}$ &  12.5$^{+7.9}_{-6.4}$\\
$\Gamma$  & 1.73$^{+0.20}_{-0.18}$ & 1.49$^{+0.21}_{-0.19}$ & 1.39$^{+0.10}_{-0.22}$&  1.67$^{+0.24}_{-0.23}$\\
E(1)      & & 6.93$^{+0.23}_{-0.24}$ & 6.82$^{+0.26}_{-0.11}$ &  \\
F(1)      & & 11.7$^{+4.4}_{-4.3}$    & 13.4$^{+5.1}_{-4.0}$ \\
E(2)      & & & 8.86$^{+0.38}_{-0.35}$ \\
F(2)      & & & 6.9$^{+3.6}_{-3.2}$ \\
Flux(2-10)& 1.52$^{+0.01}_{-0.13}$ & 1.58$^{+0.04}_{-0.21}$ & $>$1.36 &  1.51$^{+0.03}_{-0.19}$ \\
$\chi^2$  & 55.2/31=1.78 & 37.0/29=1.27 & 25.7/27=0.95 & 31.7/31=1.02\\
\tableline
&& Obs. 6 && 17\%\\
\tableline
$N_{\rm H}$ & 14.3$^{+4.1}_{-3.6}$ &  10.8$^{+4.1}_{-3.8}$ & 8.5$^{+4.7}_{-3.6}$ & 13.4$^{+4.9}_{-4.2}$\\
$\Gamma$  & 1.81$^{+0.13}_{-0.13}$ &  1.66$^{+0.14}_{-0.07}$ & 1.56$^{+0.16}_{-0.14}$ & 1.78$^{+0.16}_{-0.16}$\\
E(1)      & &  7.02$^{+0.19}_{-0.30}$ & 6.82$^{+0.18}_{-0.16}$ \\
F(1)      & &  10.5$^{+4.4}_{-4.1}$ & 13.8$^{+3.9}_{-5.0}$ \\
E(2)      & & & 8.53$^{+0.25}_{-0.27}$ \\
F(2)      & & & 8.0$^{+3.6}_{-3.3}$ & \\
Flux(2-10)& 1.94$^{+0.02}_{-0.09}$ & 1.98$^{+0.03}_{-0.12}$ & 2.05 &  1.93$^{+0.03}_{-0.16}$\\
$\chi^2$  & 57.4/31=1.85 & 41.0/29=1.41 & 26.5/27=0.98 &  31.7/31=1.02 \\
\enddata
\tablecomments{Column densities in units of 10$^{22}$ cm$^{-2}$; line
  energies in keV; line fluxes in units of 10$^{-4}$ cm$^{-2}$
  s$^{-1}$; and 2--10 keV fluxes in units of 10$^{-10}$ ergs cm$^{-2}$ s$^{-1}$}
\end{deluxetable}

\clearpage

\begin{deluxetable}{lcccc}
\tabletypesize{\scriptsize}
\tablecaption{Best-fit \emph{INTEGRAL} Spectral Parameters for Cen~A
  Using OSA~5.0\label{tab:50}}
\tablewidth{0pt}
\tablehead{
\colhead{Parameter} & \colhead{Obs. 4} & \colhead{Obs. 5} & \colhead{Obs. 6} 
& \colhead{Combined}
}
\startdata
& JEM-X & (3--30 keV) &\\
\tableline
$N_{\rm H}$\tablenotemark{a} & 12.1$^{+2.1}_{-2.2}$ & 16.9$^{+9.0}_{-7.2}$ &
15.9$^{+5.6}_{-5.0}$ & 13.4$^{+2.1}_{-2.4}$\\
$\Gamma$ & 2.10$^{+0.10}_{-0.10}$ & 1.87$^{+0.30}_{-0.26}$ &
1.87$^{+0.17}_{-0.19}$ & 1.97$^{+0.08}_{-0.10}$ \\ 
Flux(Fe)\tablenotemark{b} & $\leq7.4$ & $\leq10.6$ & $\leq6.8$ & $\leq5.8$\\
Flux(2--10)\tablenotemark{c} & 3.05$^{+0.06}_{-0.07}$ & 1.32$^{+0.02}_{-0.27}$ &
1.60$^{+0.04}_{-0.12}$ & 2.09$^{+0.02}_{-0.05}$ \\
$\chi^2$/dof & 176/160=1.10 & 176/160=1.10 & 171/160=1.07 & 201/160=1.26 \\
\tableline
& ISGRI & (17.5--200 keV)& \\
\tableline
$\Gamma$ & 1.92$^{+0.03}_{-0.02}$ & 1.93$^{+0.04}_{-0.04}$ &
1.89$^{+0.04}_{-0.03}$ & 1.92$^{+0.02}_{-0.02}$\\ 
Flux(20--100)\tablenotemark{d} & 9.35$^{+0.07}_{-0.08}$ & 5.22$^{+0.04}_{-0.08}$ &
6.20$^{+0.06}_{-0.07}$ & 6.96$^{+0.04}_{-0.06}$\\
$\chi^2$/dof & 28.8/27=1.07 & 24.3/27=0.90 & 36.2/27=1.34 & 24.4/26=0.94\\
\tableline
& SPI & (20--240 keV)&\\
\tableline
$\Gamma$ & 1.83$^{+0.11}_{-0.10}$ & 1.77$^{+0.18}_{-0.18}$ &
1.97$^{+0.17}_{-0.16}$ & 1.82$^{+0.08}_{-0.08}$ \\ 
Flux(20--100)\tablenotemark{c} & 10.6$^{+0.2}_{-0.9}$ & 6.4$^{+0.1}_{-1.3}$ &
7.2$^{+0.2}_{-1.1}$ & 8.03$^{+0.07}_{-0.43}$\\
$\chi^2$/dof & 68.7/61=1.13 & 78.1/61=1.18 & 61.1/61=1.00 &  89.0/61=1.46\\
\tableline
& \emph{INTEGRAL} & (3--240 keV)\\
\tableline
$N_{\rm H}$\tablenotemark{a} & 7.1$^{+1.0}_{-0.9}$ & 16.2$^{+3.4}_{-3.1}$ &
15.0$^{+2.5}_{-2.3}$ & 10.3$^{+1.0}_{-1.0}$\\
$\Gamma$ & 1.84$^{+0.02}_{-0.02}$ & 1.85$^{+0.04}_{-0.04}$ &
1.82$^{+0.03}_{-0.03}$ & 1.82$^{+0.02}_{-0.02}$ \\ 
Flux(Fe)\tablenotemark{b} & 9.1$^{+4.0}_{-3.6}$ & 5.7$^{+4.6}_{-5.4}$ & $\leq6.7$ & 5.6$^{+2.2}_{-2.6}$\\
Flux(2--10)\tablenotemark{c} & 3.13$^{+0.06}_{-0.06}$ & 1.33$^{+0.07}_{-0.06}$ &
1.60$^{+0.07}_{-0.03}$ & 2.12$^{+0.02}_{-0.03}$ \\
Flux(20--100)\tablenotemark{c} & 7.13$^{+0.14}_{-0.15}$ & 4.26$^{+0.18}_{-0.24}$ &
5.35$^{+0.20}_{-0.18}$ & 5.86$^{+0.11}_{-0.10}$\\
$\chi^2$/dof & 255/228=1.12 & 242/228=1.06 & 250/228=1.10 & 289/228=1.27 \\

\enddata

\tablenotetext{a}{Absorption in units of 10$^{22}$ equivalent H atoms cm$^{-2}$}
\tablenotetext{b}{Fe line flux or upper limits in units of photons cm$^{-2}$ s$^{-1}$ at 6.4 keV}
\tablenotetext{c}{Flux in units of 10$^{-10}$ ergs cm$^{-2}$ s$^{-1}$ from 2--10 keV}
\tablenotetext{d}{Flux in units of 10$^{-10}$ ergs cm$^{-2}$ s$^{-1}$ from 20--100 keV}
\end{deluxetable}

\end{document}